\documentclass{aa}  
\usepackage{graphicx}
\usepackage{txfonts}
\usepackage[]{hyperref} 
\usepackage{booktabs} 
\usepackage{array} 
\usepackage{paralist} 
\usepackage{verbatim} 
\usepackage{ulem}
\usepackage{xspace}
\usepackage{siunitx}
\usepackage{amssymb}
\usepackage{capt-of}
\usepackage{amsmath}
\usepackage{amssymb}
\usepackage{natbib}
\bibpunct{(}{)}{;}{a}{}{,} 

\graphicspath{{./img/}}
\usepackage{tablefootnote}

\makeatletter
\renewcommand*\aa@pageof{, page \thepage{} of \pageref*{LastPage}}
\makeatother

\addto\extrasenglish{%
}

\begin{document} 

\titlerunning{Indications of CO$_2$ in the atmosphere of GJ\,1214\,b from high-resolution K-band spectroscopy}
     \title{Cloudy with a chance of metals: Indications of CO$_2$ in the atmosphere of GJ\,1214\,b from high-resolution K-band spectroscopy} 
   \author{L. Nortmann \inst{1} \and
            D. Cont \inst{2,3}  \and
            F. Lesjak \inst{4} \and 
            A. D. Rains\inst{5,6} \and 
            A. Lavail\inst{7} \and 
            L. Boldt-Christmas\inst{5}\and  
            E. Nagel\inst{1} \and
            A. Reiners\inst{1} \and
            N.~Piskunov\inst{5} \and
            F. Yan \inst{8} \and    
            A. Hatzes\inst{9} \and           
            O. Kochukhov\inst{5} \and
            D. Shulyak \inst{10}\and
            U. Seemann\inst{11} \and
            M. Rengel\inst{12} \and
            A. Hahlin\inst{13}  }

   \institute{
            Institut f\"ur Astrophysik und Geophysik, Georg-August-Universit\"at, Friedrich-Hund-Platz 1, 37077 G\"ottingen, Germany \\email: Lisa.Nortmann@uni-goettingen.de\and   
            Universit\"ats-Sternwarte, Ludwig-Maximilians-Universit\"at M\"unchen, Scheinerstrasse 1, 81679 M\"unchen, Germany \and
            Exzellenzcluster Origins, Boltzmannstraße 2, 85748 Garching, Germany \and
            Leibniz Institute for Astrophysics Potsdam (AIP), An der Sternwarte 16, 14482 Potsdam, Germany \and
            Department of Physics and Astronomy, Uppsala University, Box 516, 75120 Uppsala, Sweden \and
            Instituto de Astrofísica, Pontificia Universidad Católica de Chile, Av. Vicuña Mackenna 4860, 782-0436 Macul, Santiago, Chile \and            
            Institut de Recherche en Astrophysique et Planétologie, Université de Toulouse, CNRS, IRAP/UMR 5277, 14 avenue Edouard Belin, F-31400, Toulouse, France \and
            Department of Astronomy, University of Science and Technology of China, Hefei 230026, China \and         
            Thüringer Landessternwarte Tautenburg, Sternwarte 5, 07778 Tautenburg, Germany \and       
            Instituto de Astrofísica de Andalucía - CSIC, Glorieta de la Astronomía s/n, 18008 Granada, Spain\and
            European Southern Observatory, Karl-Schwarzschild-Str. 2, 85748 Garching bei München, Germany\and
            Max-Planck-Institut für Sonnensystemforschung, Justus-von-Liebig-Weg 3, 37077 Göttingen, Germany\and
            Astrophysics Group, Keele University, Staffordshire ST5 5BG, United Kingdom}

   \date{Received 10 December 2025 / Accepted 35 March 2026}

  \abstract 
{Sub-Neptune extra-solar planets are abundant in the Milky Way, yet their atmospheric properties remain poorly understood. They frequently exhibit  muted transmission spectra, with GJ\,1214\,b being the most prominent example. Following years of  intense observing campaigns yielding featureless planetary spectra, more recent observations with JWST have revealed the first possible atmospheric signatures of H$_2$O, CH$_4$, and CO$_2$.} {We present high-resolution transmission spectroscopy of GJ\,1214\,b based on eight transits obtained with the CRIRES$^+$ spectrograph in the K band.}{We used \texttt{SYSREM} to remove telluric and stellar signals from the data and searched for signatures of H$_2$O, CO, CH$_4$, H$_2$S, NH$_3$, and CO$_2$ using the cross-correlation technique.} {We obtained non-detections for the first five molecules and used injection recovery tests to derive upper limits on the atmosphere. 
For CO$_2$ we measure a cross-correlation signal at S/N $\sim 3.6$, with a detailed investigation of the signal showing no obvious indication that it is caused by correlated noise. 
A Welch $t$-test confirmed that the in-trail, in-transit distribution is significantly different from the out-of-trail distribution at a $3.4 \sigma$ confidence.
Interpreting the data using a Bayesian retrieval framework, with multiple molecular species and free chemistry, resulted in a retrieved planet temperature of $T_\mathrm{iso}=398^{+283}_{-197}$ K,
consistent with a value intermediate between the day- and night-side temperatures from JWST-derived temperature-pressure profiles at high altitudes, as expected for the planetary terminator. In addition, a metallicity of [M/H]$ = 0.48^{+0.89}_{-1.70}$ was derived from the abundances of the retrieved molecules, along with an opacity deck pressure of  $\log_{10}(P_\mathrm{c}) = -3.04^{+2.52}_{-1.53}$.  A simpler equilibrium chemistry retrieval assuming CO$_2$ as the sole opacity source returned a compatible temperature, with smaller formal uncertainties ($T_\mathrm{iso}=509^{+102}_{-59}$ K), slightly higher metallicity ([M/H]$ =1.51^{+0.68}_{-0.75}$), and higher opacity deck pressure ($\log_{10}(P_c) = -0.88^{+1.95}_{-2.48}$). While these sets of values correspond to relatively large signal amplitudes predicted for CO$_2$ features in the mid-infrared, they are compatible with JWST NIRSpec observations within the models' $1.5\sigma$ uncertainties.}
{Further modelling and additional data are required to confirm the atmospheric signatures and obtain a comprehensive interpretation of low- and high-resolution data. Overall, our results support previous findings that CO$_2$ is likely to be a significant component of the atmosphere of GJ\,1214\,b.}


   \maketitle
%

\section{Introduction}
In recent decades, the discovery of thousands of exoplanets has revealed a remarkable diversity of planetary types. Among these, sub-Neptune planets were found to make up the largest population of detected planets in the Milky Way \citep{Borucki2010,Fulton2017}, yet they remain poorly understood. With their intermediate sizes and relatively low bulk densities, they have no counterpart in the Solar System,  as they populate the transition region between rocky planets and gas giants. Therefore, improving our understanding of these objects is essential for investigating planet formation and evolution.  However, their bulk densities, obtained from radius and mass measurements, are not sufficient to constrain their interior structure, as these values are consistent with a broad range of 
compositions \citep{RogersSeager2010,LopezFortney2014,Yang2024,Lin2025}. 
The most direct way to break this degeneracy is to characterise their atmospheres and obtain molecular abundances and mean molecular weights. While hot gaseous giant planets are highly amenable to atmospheric characterisation due to their large scale heights, studying these smaller, cooler planets remains significantly more challenging.\\
\indent Among the sub-Neptune population, GJ\,1214\,b was initially thought to be the most promising target for atmospheric characterisation. This was owed to its favourable planet-to-star radius ratio and its closeness to Earth \citep{Charbonneau2009,Cloutier2021}, allowing for observations at a high signal-to-noise ratio (S/N). 
However, early low-resolution observations of GJ\,1214\,b from the ground \citep[e.g.][]{Bean2010,Bean2011,Nascimbeni2015} and from space with the Hubble Space Telescope (HST) Wide Field Camera 3 (WFC3) and the Spitzer space telescope obtained remarkably flat spectra  \citep{Berta2012,Kreidberg2014,Desert2011,Fraine2013}. 
Similarly flat spectra have since been observed for several other sub-Neptunes, implying that this may be a common trait among the population \citep{Knutson2014,CrossfieldKreidberg2017,Benneke2019,LibbyRoberts2020,Dymont2022,Brande2024,Cadieux2024,Damiano2024,Wallack2024}. These featureless spectra can be explained by either a high-metallicity atmosphere or an optically thick cloud or haze layer in these planets' atmospheres, obscuring the molecular features below. For GJ\,1214\,b, the precision reached with 15 transits obtained by HST was sufficient to conclude that high-metallicity alone could not explain the lack of features, leading to the interpretation that it must be shrouded in high-altitude aerosols \citep{Kreidberg2014}.\\
\indent In parallel, several attempts to detect atoms in GJ\,1214\,b's atmosphere have been made with high-resolution spectroscopy from the ground. This resulted in a tentative indication for H$_\alpha$ reported by \citet{Murgas2012}, along with a detection of the metastable helium triplet at 1083 nm presented by \citet{Orell2022}. However, independent observations carried out both prior to and following the work of \citet{Orell2022} did not yield any detections of helium \citep{Crossfield2019, PddlRoche2020, Kasper2020, Spake2022, Allart2023, Masson2024}.\\
\indent Recent observations obtained with the James Webb Space Telescope (JWST) Mid-Infrared Instrument (MIRI; 4.9 to \mbox{27.9 $\mathrm{\mu}$m}) and the Near Infrared Spectrograph (NIRSpec) using grating G359H  ($\sim$  2.87 to 5.19 $\mathrm{\mu}$m) have revealed possible molecular signals for the planet.  
JWST MIRI was used to obtain a full phase curve, including the primary transit, whereby the planet passes in front of its host star in the observer's line of sight, and the secondary eclipse, when the planet disappears behind the star. The phase curve observations were used to constrain GJ\,1214\,b's bond albedo to 0.51 $\pm$ 0.06 and derive its day-side and night-side equilibrium temperatures \citep{Kempton2023}. The large phase curve variation observed for GJ\,1214\,b was best fitted with general circulation models of atmospheres with metallicities $\ge 100$ times solar.  The day- and night-side emission spectra showed signatures of absorption features, most likely attributable to water, at a $2.5\sigma$ and 2.6$\sigma$ confidence, respectively. The amplitudes of these features were compatible with abundances corresponding to metallicities of $\ge 100$ times solar, in agreement with the phase curve modelling \citep{Kempton2023}. Based on the lack of features in the MIRI transmission spectrum, \citet{Gao2023} ruled out metallicities $\le 300$ times solar metallicity. Instead, they found  that models with extremely high metallicities of $\ge1,000$ times solar metallicity or a secondary atmosphere composed primarily of gases at least as heavy as water vapour were preferred in combination with significant haze contribution. 
JWST NIRSpec observations of the transmission spectrum revealed signatures compatible with absorption by CH$_4$ and CO$_2$ at a 2.0$\sigma$ and 2.4$\sigma$ confidence, respectively \citep{Schlawin2024,Ohno2025}, although the authors note  that the methane feature might be attributable to haze absorption, rather than to CH$_4$ \citep{Ohno2025}. The modelling of the molecular features, together with different realistic hazes, led to the conclusion that the data could best be explained by an extremely high metallicity scenario ([M/H]$=3.69\pm0.16$). \citet{Ohno2025} noted that the high metallicity interpretation of GJ\,1214\,b's panchromatic low-resolution spectrum challenges our understanding of the formation and evolution of sub-Neptunes \citep{Nixon2024}. Similarly extreme values have recently been derived for some sub-Neptunes (e.g. LTT\,9779\,b, \citeauthor{Hoyer2023} \citeyear{Hoyer2023}; L98-59 d,  \citeauthor{Banerjee2024} \citeyear{Banerjee2024}) while others appear to have a more moderate enrichment of $\sim 100$ times solar metallicity (e.g. K2-18\,b, \citeauthor{Madhusudhan2023} \citeyear{Madhusudhan2023}; TOI-270\,b, \citeauthor{Benneke2024} \citeyear{Benneke2024}; and GJ\,3470\,b \citeauthor{Beatty2024ApJ} \citeyear{Beatty2024ApJ}), suggesting a diversity of atmospheric scenarios among this population \citep{Madhusudhan2025}.\\
\indent While more JWST observations are needed to confirm the molecular detections in GJ\,1214\,b's atmosphere in the mid-infrared (MIR), studies of molecular features at high-resolution from the ground in the near-infrared (NIR) may also contribute to the characterisation of this planet. At high-resolution, observations are sensitive to the resolved cores of molecular lines and, thus, to their signatures in the upper atmosphere above cloud or haze decks \citep{Snellen2010, Gandhi2020}; whereas at low-resolution, the signal of these narrow signatures would be too blended with the continuum to be detectable. High-resolution spectroscopy is also sensitive to the exact line patterns and ratios of the molecules and  can thus be used to unambiguously determine the molecule responsible for an absorption feature, avoiding confusion with broadband continuum opacity, haze, or cloud features. Such high-resolution studies have recently been conducted for the sub-Neptune planets GJ\,3470\,b \citep{Dash2024}, GJ\,436\,b \citep{Grasser2024,Pelaez-Torres2025arXiv}, and GJ\,3090\,b \citep{Parker2025}, as well as for warm Earth-sized planets such as GJ\,1132\,b \citep{Palle2025b}. While no features were detected, the observations were useful to derive upper limits on the planets' atmospheric compositions. \\
In this work, we present the results of eight transit observations of GJ\,1214,\,b obtained with the CRyogenic InfraRed Echelle Spectrograph (CRIRES$^+$) at high-resolution in the K band, a wavelength range that had previously gone unstudied for this planet. We used the data to investigate its transmission spectrum for signatures of molecules and constrain its atmospheric properties.\\
This work is structured as follows. The observations and data analysis methods are described in Sects.~\ref{sec:obs} and \ref{sec:analysis}, respectively. We present and discuss our results in Sect.~\ref{sec:resultsdiscussion}, with a special sub-section focussing on the results of the CO$_2$ analysis in Sect. \ref{sec:co2investigation}. We compare the results to space-based observations in Sect.~\ref{sec:comparetojwst}. Lastly, a summary and our conclusion are provided in Sect. \ref{sec:summaryconclusons}.

\begin{figure}[h!]
\centering
\includegraphics[width=1\hsize]{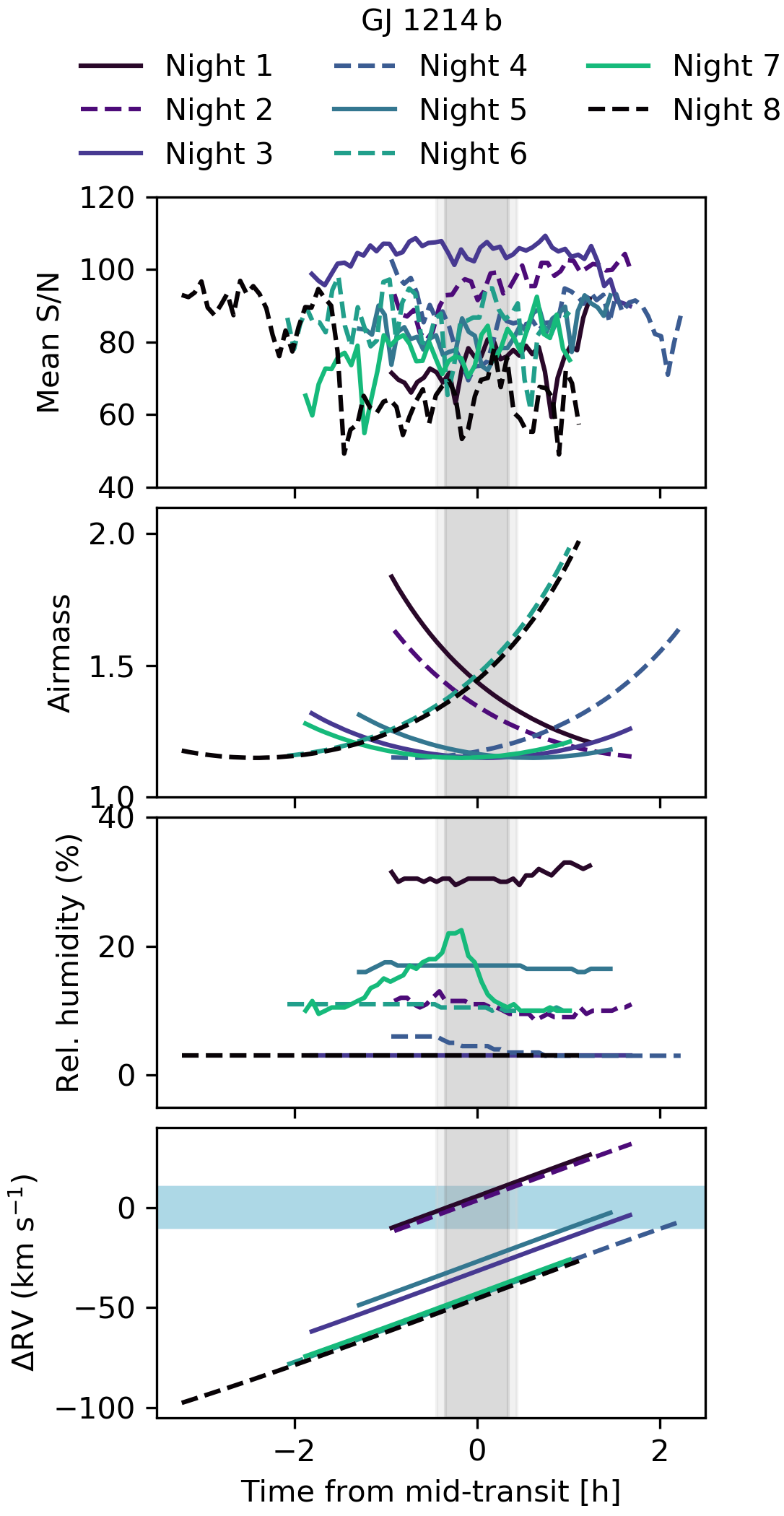}
  \caption{Conditions during the eight observations of GJ\,1214\,b. Top to bottom:\ Mean S/N per pixel for each spectrum, the progression of airmass, the relative humidity, and the absolute radial velocity shift ($\Delta$RV) between the planet and Earth rest frame are shown. The latter consist of the sum of the barycentric velocity, which changes during the observation, and the static system velocity. The transit window is indicated as a grey shaded region, with lighter shades corresponding to ingress and egress phases. The horizontal blue shaded region in the bottom panel indicates the interval with RV shifts between \mbox{$-10$ km s$^{-1}$} and \mbox{+10 km s$^{-1}$} from the telluric rest frame, e.g. a region that is especially contaminated by telluric lines.}
   \label{fig:obs_stats}
\end{figure}

\section{Observations}\label{sec:obs}
\begin{table*}\renewcommand{\arraystretch}{1.5}
\caption{Observational conditions for the eight nights of observations of GJ\,1214\,b.}  
\centering                        
\begin{tabular}{c c c c c c c c c}  
\hline\hline  
Night & Date & Duration &Exptime& \textit{N}$_\mathrm{exp}$ &Mean S/N & Airmass & Humidity  & RV shift  \\ 
 &  & (tot. / pre / post)& (s) & &&  (start - min - end) & (\%) &  (km/s)\\ 
\hline  
1 & 2022 Mar 11 & 2.2h / 0.4h / 0.7h & 240& 32 & 72 & 1.84 - 1.21 - 1.21 &31&$5.6$\\  
2 & 2022 Mar 30 & 2.5h / 0.4h / 1.2h &240&  38 & 93 & 1.63 - 1.15 - 1.15 &10&$3.5$ \\  
3 & 2022 Jul 03 & 3.4h / 1.3h / 1.2h &240&  50 & 100 & 1.32 - 1.15 - 1.26 &3&$-31.8$\\  
4 & 2022 Aug 10 & 3.1h / 0.4h / 1.7h & 240& 46 & 84 & 1.15 - 1.15 - 1.65 &4&$-44.0$\\  
5 & 2023 Jun 22 & 2.7h / 0.8h / 1.0h& 240& 40 & 81 & 1.31 - 1.15 - 1.18 &17 &$-27.1$\\  
6 & 2023 Aug 10 & 3.0h / 1.6h / 0.5h  & 240& 44 & 83 & 1.16 - 1.16 - 1.95 &11&$-44.0$ \\  
7 & 2024 Aug 06 & 2.8h / 1.4h / 0.5h & 240& 40 & 40 & 1.28 - 1.15 - 1.21 &14&$-43.0$ \\  
8 & 2024 Aug 17 & 4.3h / 2.7h / 0.6h & 240& 62 & 62 & 1.18 - 1.15 - 1.97 &3 &$-45.4$\\  
\hline  
  \label{tab:observations}
\end{tabular}
\tablefoot{Mean S/N is given per pixel. Observing duration is given in total and for post and pre transit. The absolute RV shift is given between Earth and the stellar rest frame (the latter of which is equal to the planet rest frame at the time of mid-transit).}
\end{table*}
We observed eight transits of the exoplanet GJ\,1214\,b using CRIRES$^+$, a high-spectral-resolution NIR spectrograph mounted at the Nasmyth B focus of the 8m Unit Telescope~3 (UT3) telescope at the Very Large Telescope (VLT) Facility of the European Southern Observatory (ESO) \citep{Dorn2023}. The observations were obtained between the years 2022 to 2024 as part of the CRIRES$^+$ Guaranteed Time Observations (GTO) programme (ESO Programmes:  108.22CH PI: Nagel\footnote{This transit was also part of the 108.22PH programme, but was likely misattributed to the programme 108.22CH  by the archive due to the GTO observing nights being shared between different programs.}, and 108.22PH, 109.23HN, 111.254J, 113.26GE PI: Nortmann)\footnote{The extracted data, including the refined wavelength solution of CRIRES$^+$, are available at 
\url{https://zenodo.org/records/19387252} \citep{lavail_2026_19387252}.}. 
All eight observations covered the full transit event and also before and after transit baseline in each night. The observations each night spanned between 2.2h and 4.3h (an average of 3h). We set the slit width to 0.2” and used the adaptive optics mode to achieve the nominally highest resolution of CRIRES$^+$ ($\mathcal{R}\sim$100,000). This resolution translates into a velocity resolution of approximately 3 km s$^{-1}$, being sampled onto 3 pixels.  The observations were obtained in nodding mode using an ABBA nodding pattern. This means that the telescope was moved between two positions on the sky, moving the star 6'' along the slit between the two positions (A and B, respectively), which was used in Sect. \ref{sec:preproc} to remove the sky background flux. We chose the K-band setting K2148 that covers the wavelength range between 1972 - 2452 nm, with wavelength gaps, in six echelle orders for both the A and B positions on the slit. Each echelle order is split over the three detectors of the instrument, dividing our total covered wavelength range into 18 separate segments. 
The exposure time per frame was 240s (DIT=240s, NDIT=1). The planet radial velocity changed about 1 km s$^{-1}$ during a single exposure (i.e. by 1 pixel on the detector) and from $-7.5$ to +7.5 km s$^{-1}$ with respect to the barycentre of the GJ\,1214 system over the duration of the transit events. In total, 352 spectra were taken over the eight nights, with 96 of them taken during transit. The dates and durations of the observations, the average S/N per pixel 
and the number of spectra taken are summarised in Table \ref{tab:observations}, together with other observing parameters.  The progression of the S/N and airmass over the individual observing nights is shown in Fig.~\ref{fig:obs_stats}. The data reduction and determination of the wavelength solution are described in Sect \ref{sec:preproc}.

\section{Analysis}\label{sec:analysis}
\begin{figure}[h!]
\centering
\includegraphics[width=1\hsize]{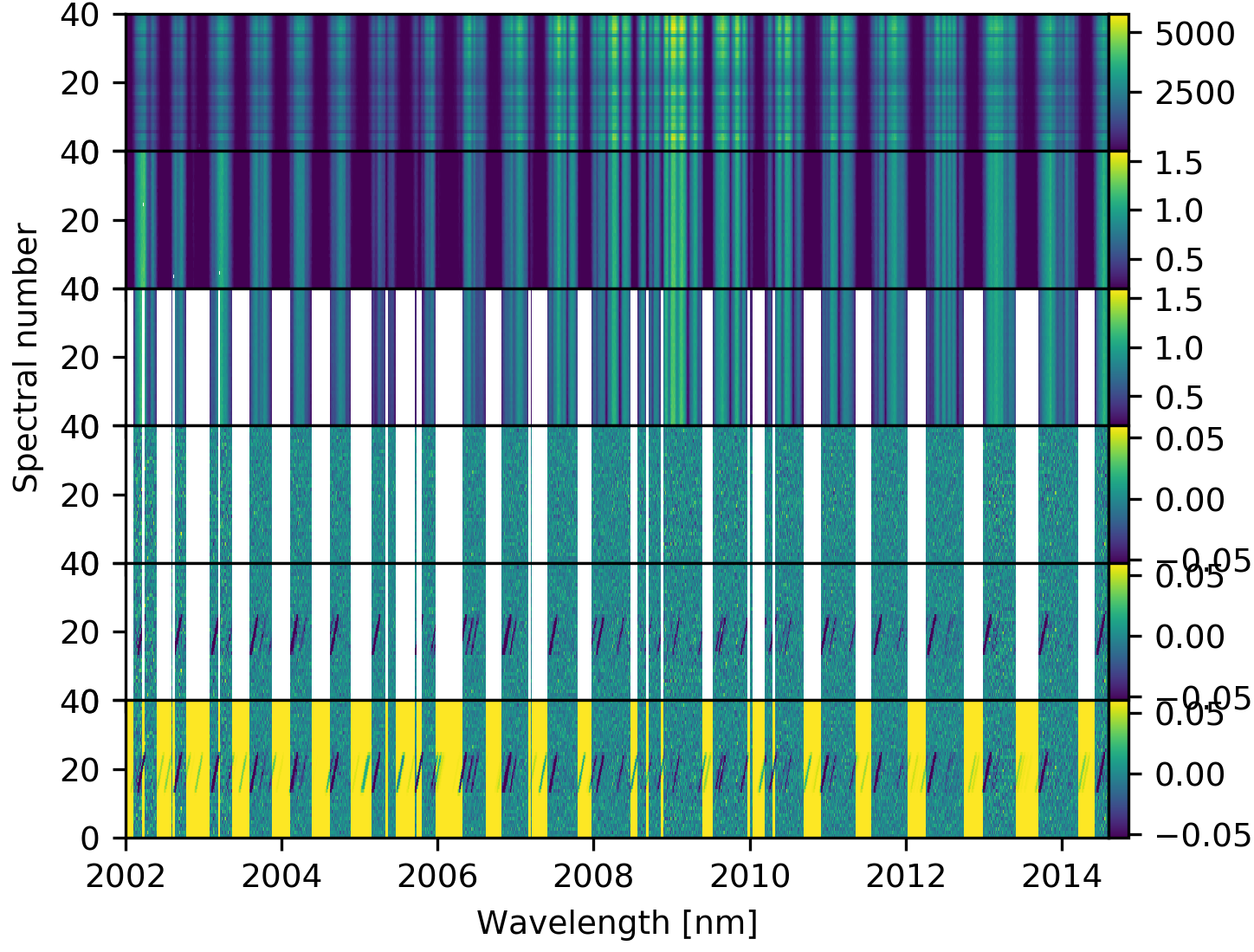}
  \caption{Stages of the data after progressive analysis steps, shown on the example of the third wavelength segment (2002.0-2014.6 nm) of night 5. Top to bottom:\ Data are shown as 1. the unnormalised time series; 2. after normalisation; 3. after masking of the deepest telluric lines ($\le 20\%$ of the continuum level);  4. after nine \texttt{SYSREM} iterations. The lowest two panels show the same stage of analysis as the fourth; 5. with a CO$_2$ containing synthetic planet atmosphere injected at 200 times the strength expected for GJ\,1214\,b; and 6. with the masked regions set to zero (shown in yellow), to allow a visualisation of the amount of planet signal falling into the masked regions.}
   \label{fig:analysis}
\end{figure}

The observed spectra contained signals of Earth's atmosphere (telluric lines), the star, and the potential planetary transmission signal. The latter would only be present in the spectra obtained during transit.  To investigate the data for the presence of molecular signatures of GJ\,1214\,b, we first had to remove the stellar and telluric lines. Afterwards, we used the methodology for the high-resolution cross-correlation spectroscopy (HRCCS) method established in the literature  \citep{Snellen2010,Brogi2012,deKok2013,Birkby2013}. All analysis steps are detailed in \citet{Nortmann2025}, which we briefly summarise here. In Fig.~\ref{fig:analysis}, we show the different steps of the analysis for one wavelength segment for night 5.

\subsection{Normalisation}
We treated all A and B spectra as a single data set and normalised all spectra to a common blaze function. This was done by fitting a second-order polynomial to the ratio of each individual spectrum to a master spectrum, which was obtained by summing all out of transit spectra. In this fit, we masked all significant telluric lines, identified using a theoretical telluric spectrum generated by \texttt{Molecfit}. Masked were all regions that in this theoretical telluric spectrum fell below 96\% of the continuum level. This mask was only applied for the normalisation step. A weaker mask, targeting only the cores of the deepest telluric lines, was applied later in the analysis and is described further below. We performed the normalisation step separately for each wavelength segment. We also flagged pixels with ‘not a number’ (NaN) values as well as significant outliers (more than 4 $\sigma$), indicating bad or hot pixels, in this step. Dividing the individual spectra by the fitted function brought them to a common blaze function. In a second step, we then fitted the common blaze function by fitting the master spectrum with a second-order polynomial for each wavelength segment. In this step, we masked not only the telluric lines, but also the stellar lines, which were identified and masked in an iterative approach rejecting strong negative outlier points and their surrounding pixels in five iteration steps. We then divided all spectra by the fitted common blaze function. 
In a final step, we checked column by column how many pixels had been flagged as either NaNs or bad pixels and masked the entire column if more than three pixels per column were affected. Otherwise, we corrected them by linearly interpolating over the flagged values. 
Before removing telluric and stellar signals with \texttt{SYSREM} (as described in  Sect. \ref{sec:sysrem}), we masked the deepest telluric lines, specifically the regions where the flux fell below 20\% of the continuum level. This was done because the variation of flux over time is not measured accurately any more once saturation is reached. Therefore, the behaviour in the deepest lines would not follow those of weaker telluric lines. As a consequence, \texttt{SYSREM} would not be able to correct these features reliably. Furthermore the S/N in these pixels is very low. In total, about 7-10\% of the pixels were masked, varying slightly between the nights.

\subsection{Removal of telluric and stellar signals}\label{sec:sysrem}
We removed the stellar and telluric signals by using the detrending algorithm \texttt{SYSREM} \citep{Tamuz2005,Birkby2013},
which has long been established as a standard tool for high-resolution spectroscopic analysis \citep[e.g.][]{Birkby2013,Birkby2017,AlonsoFloriano2019,SanchezLopez2019,Gibson2020,Gibson2022,Czesla2024}. The algorithm makes use of the fact that stellar and telluric lines are almost static in wavelength while the planet lines move about 1 km s$^{-1}$ during each exposure, due to the change in the planet's radial velocity during the observations. Before running \texttt{SYSREM} we aligned the spectra obtained in nodding positions A and B to the same wavelength grid via interpolation. We also aligned the spectra into stellar rest frame in the same step by Doppler shifting the spectra by the stellar radial velocity, $v_\mathrm{\star} \left( t \right)$,
\begin{equation}
    \label{equation:stellar-rest-frame} 
    v_\mathrm{\star} \left( t \right) = v_\mathrm{sys} + v_\mathrm{bary} \left( t \right) + K_\mathrm{\star} \sin{2\pi\phi \left( t \right)},
\end{equation}
where $\phi$ is the phase of the orbit, $K_\mathrm{\star}$ is the stellar orbital velocity semi-amplitude, $t$ is the time of the observed spectrum, $v_\mathrm{bary}$ is the barycentric velocity, and $v_\mathrm{sys}$ is the velocity of the GJ\,1214 system with respect to the barycentre of the Solar System (see Table \ref{tab:gj1214parameters}. The first and last 21 pixels of each segment were removed from consecutive analysis to avoid the edge effects from interpolation during spectral extraction and the alignment made in this step.\\
The shift to stellar rest frame ensured that the stellar lines were fully aligned and thus could be optimally removed by \texttt{SYSREM}. We considered this step important to avoid spurious signals caused by stellar line residuals due to the fact that the host star GJ\,1214\,b is an M-dwarf and contains prominent water lines in its spectrum which overlap with the planet spectrum at mid-transit for all nights. We ran \texttt{SYSREM} for each night independently for 14 iterations, saving the output after each step. Following \cite{Gibson2020}, we divided the normalised uncorrected data by the summed-up correction matrices of each iteration to achieve the final correction of the data. \\
\indent After using \texttt{SYSREM} successfully, the data should only contain noise and the potential planet transmission spectrum. We evaluated the number of \texttt{SYSREM} iterations required to sufficiently clean the data through the two commonly used approaches of (I) monitoring the behaviour of the standard deviation of the data from one \texttt{SYSREM} iteration to the next \citep[e.g.][]{Herman2020, Herman2022} and (II) monitoring the progression of the S/N level at which we are able to recover an injected planet signal \citep[e.g.][]{Cheverall2023} as well as through the behaviour of CCF signals in our data analysis. Details can be found in Sect. \ref{Sec:apendixsysremiterations}. Based on this evaluation, we adopted the data corrected with the accumulated correction terms up to the ninth \texttt{SYSREM} iteration to proceed with our analysis.

\subsection{Synthetic model calculation}\label{sec:modelcalculation}
We investigated the residual data for molecular signals in the planet atmosphere using cross-correlation with synthetic spectra.  We searched for the molecules water (H$_2$O), methane (CH$_4$), carbon monoxide (CO), carbon dioxide (CO$_2$), hydrogen sulphide (H$_2$S), and ammonia (NH$_3$), which are expected to have appreciable signatures in the K band for the temperatures expected for  GJ\,1214\,b.\\
\indent The model spectra were calculated using \texttt{petitRADTRANS} \citep{Molliere2019}. For the temperature-pressure ($T$-$p$) profile, we set up a pressure grid spanning from $10^{-8}$ to $10^2$ bar in 125 logarithmically spaced steps and calculated the temperatures as an isothermal profile with $T_\mathrm{iso}=500$ K, which lies centred between the observed equilibrium temperatures obtained for the planet's day and night sides (\mbox{$T_\mathrm{day~side}=553$ K} and \mbox{$T_\mathrm{night~side}=437$ K}, \citeauthor{Kempton2023} \citeyear{Kempton2023}). We chose an isothermal profile as our calculated transmission spectra are not sensitive to the shape of the \mbox{$T$-$p$ profile} at higher pressures. We verified this by comparing the models calculated with the isothermal profile to those that were calculated with an \mbox{alternative $T$-$p$ profile}, with a physically expected increase of temperatures at higher pressures, and found them to be almost identical. For the comparison of the \mbox{$T$-$p$ profile}, we used Eq. 29 in \cite{Guillot2010} and assumed the values $\kappa_\mathrm{IR}=0.01$ for the atmospheric opacity in the  IR, $\gamma=0.4$ for the ratio between the optical and IR opacity, both chosen to create a moderate temperature increase at higher pressures, an interior temperature of $T_\mathrm{int}=200$ K, and an equilibrium temperature of $T_\mathrm{eq}=550$ K, the latter corresponding to the planet's day-side equilibrium temperature. The resulting profiles are shown in Fig.~\ref{fig:template_tpprofile_masfunct}.\\ 
\indent The pressure-dependent chemical abundances of each molecule and hydrogen and helium as well as the pressure-dependent mean molecular weight (MMW) were calculated based on the $T$-$p$ profile, C/O ratio, and metallicity [M/H] using a chemical grid calculated with \texttt{easyCHEM} \citep{LeiMolliere2025}. We created different template spectra for each molecule, by including only the line opacities of that molecule as well as the continuum absorption of collision-induced absorption (CIA) of H$_2$-H$_2$ and H$_2$-He.  
The molecular opacities were based on the line lists of \citet{Rothman2010} for H$_2$O, CO, and CO$_2$, \citet{Hargreaves2020} for CH$_4$, \citet{Rothman2013} for H$_2$S, and \citet{Yurchenko2011} for NH$_3$. The impact of clouds or aerosols was modelled as a fully opaque grey cloud deck located at a pressure level, $P_\mathrm{c}$. The metallicity and opacity deck pressure level were varied for different models. For the cross-correlation analysis we fixed the C/O ratio to solar values, that is, C/O=0.55 for all models.
The effects of varying metallicity and cloud deck height are illustrated in Fig.~\ref{fig:effects_on_models}.\\ 
\indent Using further input of the planet radius, $R_\mathrm{p}$, at reference pressure, $P,$ (we chose $P=0.01$ bar) and surface gravity, $g_\mathrm{p}$, \texttt{petitRADTRANS} calculates a wavelength dependent planet radius, $R_\mathrm{p,\lambda}$. We transformed this output into a transmission spectrum,  $1-\delta (\lambda)=1-(R_\mathrm{p,\lambda}/ R_\mathrm{\star})^2$, where $\delta (\lambda)$ is the transit depth. The planet and system parameters that we used are summarised in Table \ref{tab:gj1214parameters}. Finally, the transmission spectra were then convolved to match the nominal resolution of CRIRES$^{+}$, $\mathcal{R}=$100,000, using a Gaussian kernel.

\subsection{Cross-correlation search for molecular signatures} \label{sec:ccf}
We cross-correlated our residual spectra with template spectra for H$_2$O, CH$_4$, CO, CO$_2$, H$_2$S, and NH$_3$ to search for signals of these molecules. For this step, we chose templates that did not contain clouds, with a mean molecular weight and molecular abundances corresponding to a metallicity of [M/H] = 1.5, that is, about 30 times solar metallicity. The template spectra for each of the six molecules are shown in the left column of Fig.~\ref{fig:model-ccf-kp}.\\
We conducted the analysis for each wavelength segment separately and normalised the template model  spectrum for each segment by truncating the full spectrum to the wavelength region of the investigated wavelength segment before dividing the truncated model segment by its mean. We then used the weighted cross-correlation function \citep[CCF;][]{Gibson2020,Cont2022}, expressed as\begin{equation}
    \mathrm{CCF}(v, t) = \sum _{i = 0}^N \frac{R_i(t) \cdot M_i(v)}{\sigma_{i}(t)^2}\,,
\end{equation}
where $t$ and $i$ represent the time and pixel indices respectively, $R$ represents the residual spectra, $\sigma$ their corresponding uncertainties, and $M$ is the synthetic model template Doppler shifted by a velocity, $v$. The uncertainties for each pixel are taken from the pipeline and propagated through all analysis steps by dividing them by all terms applied to the data and shifting them when rest-frame corrections were performed. We explored the range of $v=-150$ km s$^{-1}$ to  $v=+150$ km s$^{-1}$ in 1 km s$^{-1}$ steps. We then summed up the CCFs of our 18 wavelength segments. 

\subsubsection{$K_\mathrm{p}$ - $v$ maps}
We brought the summed up CCFs from the stellar restframe into the planet restframe by Doppler shifting them by
\begin{equation}
    \label{equation:planetary-rest-frame} 
    v_\mathrm{p}^{\star} \left( t \right) = K_\mathrm{p} \sin{2\pi\phi \left( t \right)} - v_\star,
\end{equation}
where $K_\mathrm{p}=102.5\pm2$  km s$^{-1}$ is the planet's radial velocity semi-amplitude. We then summed up the shifted CCF along the time axis between the times of second and third contact of the transit, that is, the exposures at which the planet was fully in front of the host star.  
The CCFs of all exposures over the eight nights sorted by orbital phase of the planet and aligned to the planet rest frame are shown in the middle column of Fig.~\ref{fig:model-ccf-kp}.\\
\indent We repeated the process of alignment of the CCFs for different values of  $K_\mathrm{p}$ spanning a grid from 0 km s$^{-1}$ to  400 km s$^{-1}$ in \mbox{1 km s$^{-1}$} steps resulting in a so-called $K_\mathrm{p}$ - $v$ map. We normalised the CCF signals in this map by the standard deviation of the regions of the map where no signal was expected (defined here as the regions of the velocity ranges $v=[-150,-10]$ km s$^{-1}$ and $v=[+10,+150]$ km s$^{-1}$, each spanning the entire $K_\mathrm{p}$ range $[0,400]$ km s$^{-1}$) and dividing the map by this value. The resulting normalised maps, showing the detection strengths in units of S/N are shown in the right column of Fig.~\ref{fig:model-ccf-kp}. The planet signal is expected at  $K_\mathrm{p}=102.5\pm2$ km s$^{-1}$, calculated  based on the system parameters, and at $v=0$ km s$^{-1}$ , that is, with no relative velocity shift to the planet rest frame.

\subsection{Exploring dependence of CCF results on model template}\label{sec:modeldependence}
\begin{figure*}
\centering
\includegraphics[width=1\hsize]{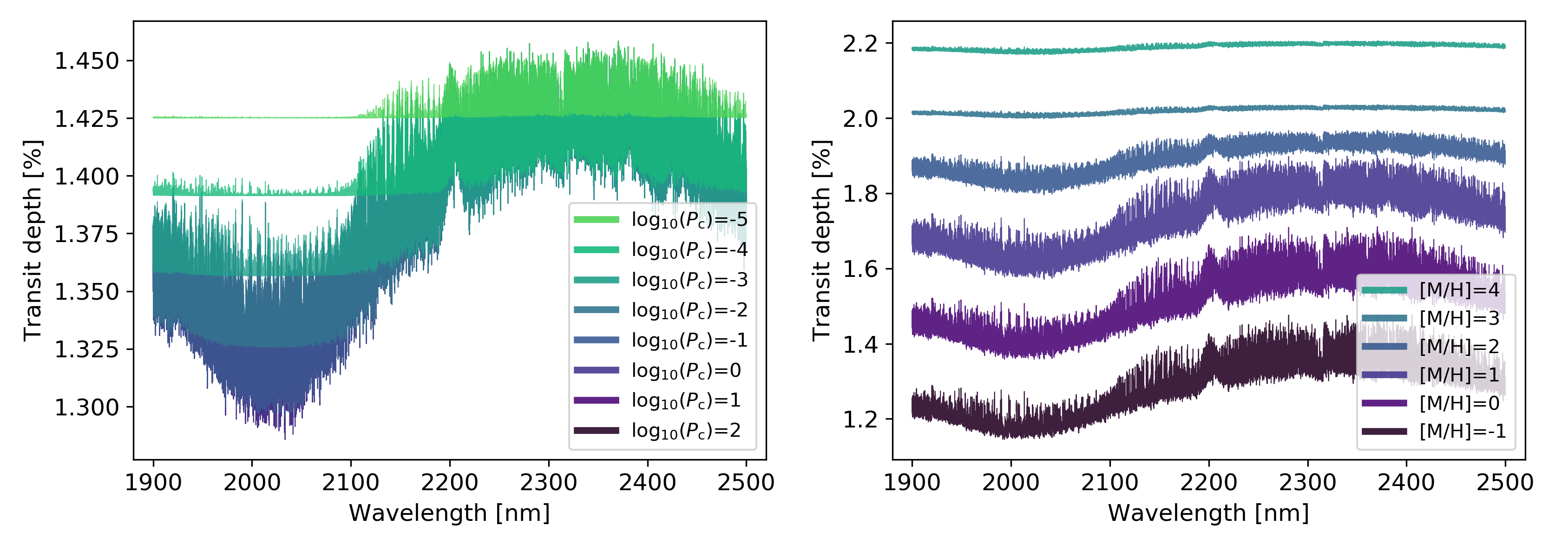}
  \caption{Effect of cloud deck height and atmospheric metallicity on planetary transmission spectra. Left panel:  Effect of increasing height of the cloud deck, by decreasing the grey cloud deck pressure, $P_\mathrm{c}$, of the models is illustrated on a model containing only CH$_4$ lines. Right panel:  Effect of decreasing atmospheric scale height, caused by increasing atmospheric metallicity [M/H], and thus the mean molecular weight, is shown with the same model without clouds.}
   \label{fig:effects_on_models}
\end{figure*}
It has been shown that if the planet signal is affected by clouds, using a template spectrum that  contains clouds at the correct pressure level yields a stronger detection in the CCF than using a cloud-free model \citep{BoldtChristmas2025}. This makes sense as the correct model will have a slightly better match to the actual absorption lines contained in the data as well as to their relative line ratios. Consequently, it could be possible that a marginal signal at the edge of detectability is overlooked when the investigation is only conducted with one fixed model (e.g. only a cloud-free model). Therefore, we repeated our analysis with a grid of template spectra spanning a wide range of metallicities and cloud deck pressures. The grid covered template spectra with  metallicities from [M/H] $ = -1.00$ to $+4.00$ in 21 steps and cloud top pressures between $P_\mathrm{c}=10^{1.7}$ and $10^{-5.5}$ bar in 19 logarithmically spaced steps. 
For each model, the mean molecular weight and molecular abundance were calculated based on the sampled metallicity and assuming a solar C/O ratio. The resulting volume mixing ratios for the individual molecules depending on the metallicity are shown in Fig.~\ref{fig:template_tpprofile_masfunct} for a subsample of the metallicities probed in the grid. We explored only one molecule at a time, that is, the models contained only the lines of that molecule and CIA continuum absorption.  The detection significance of each explored model was evaluated as the maximal S/N of any signal found in the   normalised \mbox{$K_\mathrm{p}$ - $v$}  map within the $\pm 10 $ km s$^{-1}$ vicinity of the expected position  ($K_\mathrm{p}=102.5$ $\pm 10$ km s$^{-1}$ and $v=0$ $\pm 10$ km s$^{-1}$).  The results of this analysis are shown in in Fig.~\ref{fig:detectiongrid}. 

\subsection{Injection recovery tests}\label{sec:injrecov}
We explored the sensitivity of our data set to different atmospheric scenarios by injecting synthetic model spectra into our data and evaluating at which S/N we were able to recover them. With increasing metallicity, the mean molecular weight of the atmosphere increases, leading to a smaller atmospheric scale height and a general decrease in the amplitude of absorption features, which would make them harder to detect in our data. Similarly increasing the height (decreasing the pressure) of the opacity deck (due to clouds or aerosols) can decrease the amplitude of the observable absorption lines and may even hide some of the lines beneath the opacity deck (see. Fig.~\ref{fig:effects_on_models}). \\
\indent To test our data for its sensitivity to different models we explored the same model grid used in the previous section (Sect.~\ref{sec:modeldependence}), spanning the intervals [M/H] = [$-1.00$ , $4.00$] and $P_\mathrm{c}=[10^{1.7}, 10^{-5.5}]$. We injected the model into the data before the normalisation step using the planet velocity based on literature values and repeated the analysis steps described in previous sections. We used the same model for the respective cross-correlation template that had been injected into the data for every tested model. 
The normalised $K_\mathrm{p}$ - $v$ maps were evaluated at the position $K_\mathrm{p}=102.5$ km~s$^{-1}$ and $v=0$ km~s$^{-1}$. To avoid any bias from peaks or dips in the CCF caused by the underlying real data, the $K_\mathrm{p}$ - $v$ maps stemming from the analysis of the real data with the same model template were subtracted from the  \mbox{$K_\mathrm{p}$ - $v$ maps} of the data with injected planet signal before evaluation. The results of this analysis are discussed in Sect. \ref{sec:snrgrids}.

\subsection{Bayesian retrieval}
We used a Bayesian retrieval framework to investigate a potential planetary signal in our data. For this purpose, we tried two different approaches:\ 
a simplified equilibrium chemistry retrieval with CO$_2$ and CIA absorption as the only opacity sources and a free-chemistry retrieval, including the opacities of all six molecules investigated in the cross-correlation study. For the equilibrium chemistry retrieval the framework was set up with the same basic structure for the atmospheric models as described in Sect. \ref{sec:modelcalculation}. The models only contained lines of CO$_2$ and CIA continuum absorption by hydrogen and helium. 
We used the same pressure grid as in \ref{sec:modelcalculation} and calculated the $T$-$p$-profile as isothermal with $T_\mathrm{iso}$ as a free parameter. Further free parameters of the retrieval were the C/O ratio, the metallicity ([M/H]),  the pressure ($\log_{10}{P_\mathrm{c}}$)  of a grey opacity layer (due to clouds or aerosols), the planet's radial velocity (RV) semi-amplitude, $K_\mathrm{p}$,  a constant velocity offset from the planet rest frame, $v_0$, as well as a scaling parameter, $\beta$, allowing the errors of the data to be scaled in case the errors provided by the reduction pipeline were under- or overestimated.\\
\indent For the free chemistry retrieval we fixed the orbital parameters $K_\mathrm{p}$ and $v_0$  to the values obtained in the CO$_2$-only, equilibrium chemistry retrieval. This was done to explicitly explore the possibility of other molecular signals or their absence coexisting with the CO$_2$ signal. We further exchanged the parameters for the C/O ratio and metallicity for the volume mixing ratios (VMRs) of the six molecules probed in the injection recovery test, H$_2$O, CH$_4$, CO, CO$_2$, H$_2$S, and NH$_3$ as well as that of one additional proxy molecule. The VMRs were assumed to be constant with pressure. Since the mean molecular weight plays an important role in regulating the detectability of species in planets that may host a high metallicity atmosphere, we were concerned that calculating it only based on abundances of the molecules whose opacities were included in the retrieval may lead to its underestimation.  Instead of solving this by leaving the mean molecular weight as a free parameter of the retrieval, which runs the risk of unphysical parameter combinations, we included the VMR of a seventh molecule as a free parameter. We set this molecule to be N$_2$, as we required it to have a concrete molar weight for the calculations. It was, however, meant to serve as a proxy for all species who do not have any observable features in the investigated range, but could still contribute to the mean molecular weight, rather than only N$_2$. The remainder of the atmosphere was filled with hydrogen and helium at the solar abundance ratio.\\
\indent The priors for all parameters were chosen as uniform and their limits are listed in Table \ref{tab:rerivalresults}.  The retrieval used the Markov Chain Monte Carlo method (MCMC), as implemented by \cite{Foreman-Mackey2013},  to sample the parameter space and optimise the log-likelihood. At each walker step a model was calculated with the parameter sample of the step. The model was then shifted into the rest frame of the data for each of the eight data sets and, for every exposure, weighted by the fraction of the planetary disk occulting the host star at the time. To reproduce the distortions of the planet signal introduced by using \texttt{SYSREM} on the real data, we then followed the pre-processing steps described by \citet{Gibson2022}. 
The general idea behind their approach is that \texttt{SYSREM} identifies dominant patterns in the data as time-dependent vectors and the weights with which these patterns appear in each wavelength channel. The time-dependent vectors from each \texttt{SYSREM} iteration are saved, and it is assumed that these vectors will not change when small variations in the model are introduced. Therefore, when filtering the model during the retrieval, the previously derived time vectors can be reused and only the corresponding wavelength-dependent weights need to be recalculated. These weights can be obtained through a linear fit to the data, which can be written as a simple matrix multiplication. We give more details on the equations used to facilitate these steps in this work Sect. \ref{sec:modelfiltering}.\\
The filtered model was then used to calculate the log-likelihood,
\begin{align}
\ln(L) = - \frac{1}{2}\sum_{i,j}\left(\frac{(R_{i,j} - M'_{i,j})^2}{(\beta \sigma_{i,j})^2} + \ln(2\pi(\beta \sigma_{i,j})^2)\right)\,,
\end{align}
where $R_{i,j}$ are the residual spectra at pixel $i$ and time $j$ with propagated uncertainties $\sigma_{i,j}$, $M'_{i,j}$ represents the two-dimensional (2D) matrix of the filtered model spectrum, and $\beta$ is the scaling factor for the uncertainties. We ran the retrieval with 20 walkers for 15000 steps. 
We found the chains converged quickly and discarded the first 500 steps of each walker as burn-in, using the remaining
14,500 steps (corresponding to 290,000 sampled points) to calculate the resulting probability distributions, their median values and their 1$\sigma$ confidence intervals for each of the free parameters. The results of the retrieval are discussed in Sect. \ref{sec:retrieval_results_discussion}.

\subsection{Simulation of future observations}
We simulated the expected increase of detectability of features that could  be achieved by additional observations. First, we simulated the benefit of adding another eight CRIRES$^+$ transits by using the same data sets twice in the injection-recovery analysis described in Sect. \ref{sec:injrecov} but expanded our sampled grid to metallicities up to [M/H]=5.5, in anticipation of higher sensitivity in these regions with the simulated data. For the second use of each data set we inverted the time axis to inject the model into different spectra of the time series and inverted the value of the RV semi-amplitude, so that the signal was injected at planet velocities calculated with $K_\mathrm{p,sim}=-K_\mathrm{p}$. This way the planet signal still overlapped with the stellar lines at mid-transit in the same way a true signal would, but we did not sample the same noise structure of the data set twice.\\
\indent To complement this test, we further simulated the expected constrains on the planet atmosphere as if observed with  a hypothetical K-band spectrograph mounted at ESO's Extremely Large Telescope (ELT) currently under construction in Chile. The details on the simulations of the data are found in Sect. \ref{sec:ELTsims}.

\section{Results and discussion}\label{sec:resultsdiscussion}
\begin{figure*}[h!]
\centering
\includegraphics[width=0.99\hsize]{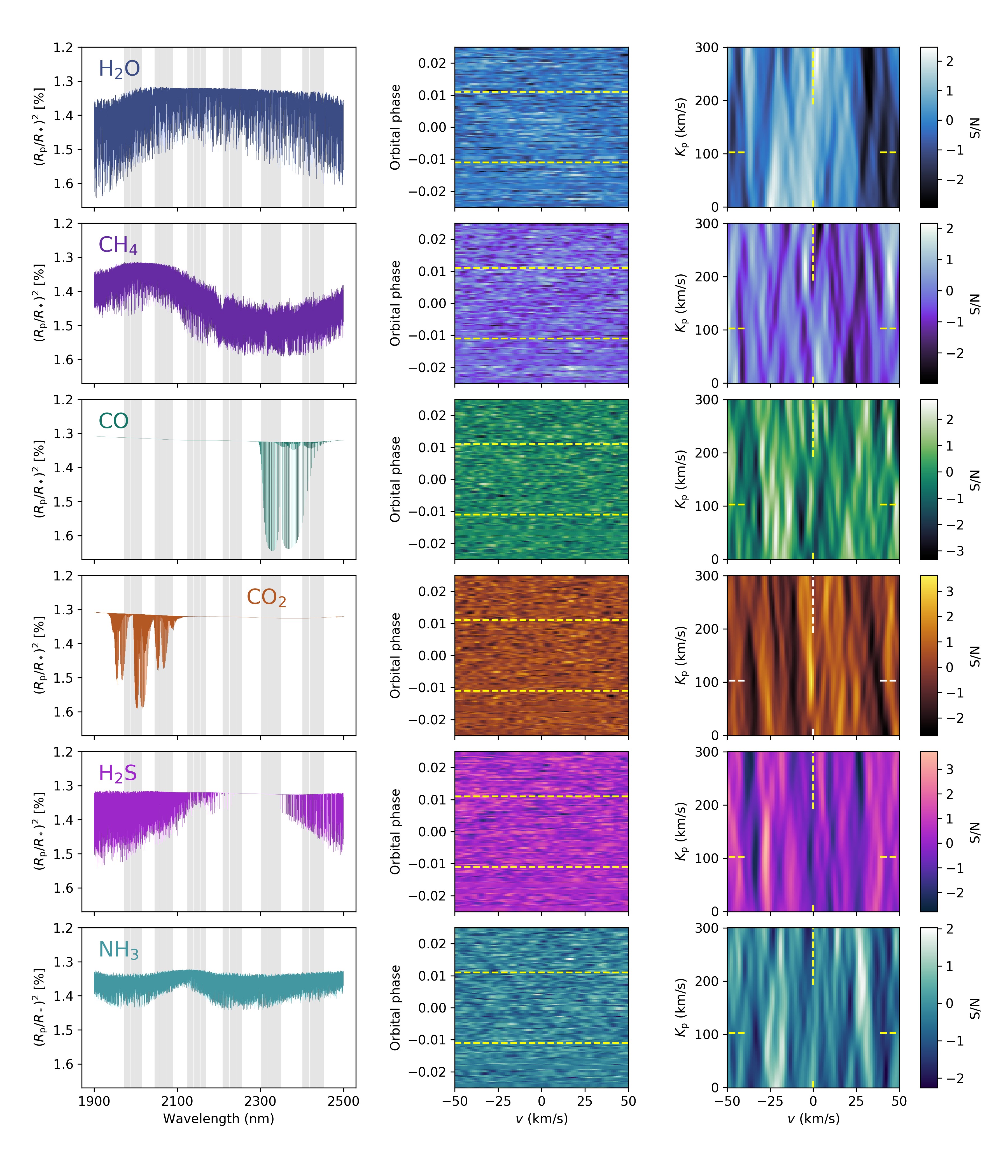}

  \caption{Results of the cross-correlation analysis. Left column: Synthetic model templates used for the individual molecules. Middle column:  Cross-correlation functions for every spectrum of all eight nights, sorted in phase and shifted to planet rest frame. The expected position of a signal should be vertical at {$v=0~\mathrm{km~s}^{-1}$} between the two horizontal dashed lines which indicate the first and last point of contact of the transit. Right column: $K_\mathrm{p}-v$ maps for each molecule. The expected position of a planetary signal is indicated by the dashed lines at $K_\mathrm{p}$=102.5 km s$^{-1}$ and $v=0$ km s$^{-1}$.}
     \label{fig:model-ccf-kp}
\end{figure*}
\subsection{Searches for molecular signals}
We investigated the high-resolution transmission spectrum of GJ\,1214\,b for signals of
H$_2$O, CH$_4$, CO, CO$_2$, H$_2$S, and NH$_3$. 
For this purpose, we cross-correlated the data with synthetic spectra (left column of Fig.~\ref{fig:model-ccf-kp}). The middle column of Fig.~\ref{fig:model-ccf-kp} presents the CCFs of each of the 384 exposures across the eight nights, sorted by orbital phase and shifted into the planetary rest frame, while the right column shows the resulting  $K_\mathrm{p}$ - $v$ maps. 
We did not find statistically significant signals for any of the investigated molecules, but we observed a CCF signal just below our significance threshold for CO$_2$. In this context, we defined a significant signal as a peak in the $K_\mathrm{p}$ - $v$ map with an amplitude exceeding S/N = 4. While there is no consensus on a precise cut-off value, a threshold of \mbox{S/N = 4 to 5} is commonly adopted before assuming a  detection to be robust, as peaks below this level can still arise from correlations with the noise \citep[e.g.][]{Schwarz2015SNR,Gandhi2020SNR34}. Signals around and below the threshold need to be evaluated carefully, for example by employing further statistical tests. \\
For H$_2$O, CH$_4$, CO, H$_2$S, and NH$_3$ the maximum CCF signals in the region around the expected planetary signal (here, defined as $K_\mathrm{p}=102.5 \pm 10\mathrm{~km~s}^{-1}$ and $v_0=0\pm 10\mathrm{~km~s}^{-1}$) remain below S/N = 2.2. This is still the case when exploring a wide range of metallicities and opacity deck pressures for the model templates underlying the analysis as shown in Fig.~\ref{fig:detectiongrid}. However, for CO$_2$ we observed a signal at S/N = 3.6, peaking at $K_\mathrm{p,CO_2}=101\mathrm{~km~s}^{-1}$ and $v_\mathrm{CO_2}=-1 \mathrm{~km~s}^{-1}$, and being the highest signal peak in the entire $K_\mathrm{p}$ - $v$ map. Its good agreement with the expected planet velocity makes this signal a candidate planet atmosphere signal which would benefit from additional observations to exceed the S/N = 4 threshold. We performed a Welch $t$-test comparing the population of CCF values inside the planet trail with those outside. The test shows that the in-trail distribution differs from the out-of-trail distribution at a 3.4$\sigma$ significance level (see Sect.~\ref{sec:inouttrail} for details). 
A more detailed investigation of this signal is presented in Sect. \ref{sec:co2investigation}. Table \ref{tab:CCFpeaks} summarises the positions and amplitudes of peaks in the $K_\mathrm{p}$ - $v$ maps for three different cases: in the region around the expected planetary signal, at the exact expected planetary signal position, and at the position at which the CO$_2$ signal peak was found.

\begin{figure*}
\centering
\includegraphics[width=1\hsize]{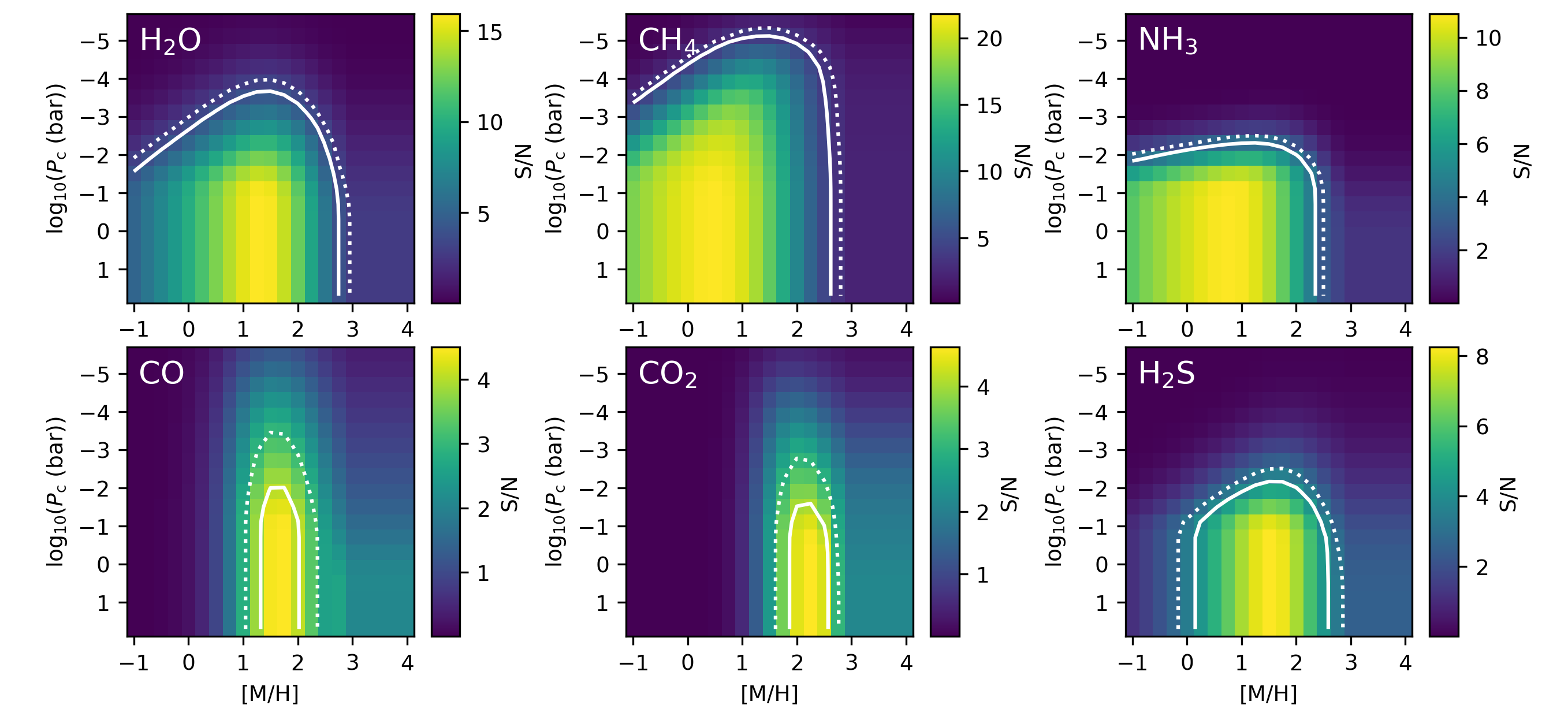}
  \caption{Results of the S/N grid analysis that show the recovered detection significance of injected synthetic models with varying metallicities [M/H] and opacity deck pressures, $P_\mathrm{c}$. The injected models contained CIA continuum absorption and lines of one molecule (H$_2$O, CH$_4$, CO, CO$_2$, H$_2$S, and NH$_3$, respectively). The white lines represent the  contours of S/N=3 (dotted) and S/N=4 (solid).}
     \label{fig:snrgrid}
\end{figure*}

\subsection{Upper limits from injection recovery test}\label{sec:snrgrids}
As we obtain clear non-detections for most of the probed molecules, we can use these to derive upper limits on GJ\,1214\,b's atmospheric conditions. For this we explored the sensitivity of our eight-transit data set to the signals of atmospheric scenarios spanning a range of different metallicities and opacity deck pressures. This was done by injecting these models into the data and determining at which S/N we were able to recover them. Regions of the parameter space in which we would have been able to achieve a detection of S/N=3 to 4  or higher can thus be excluded for the molecules H$_2$O, CH$_4$, CO, H$_2$S, and NH$_3$ for which we did not achieve such a detection in the real data. The results are shown in Fig.~\ref{fig:snrgrid}. \\
\indent If we apply the S/N = 4 threshold, we find that our sensitivity threshold at high metallicities appears to be very similar across all species, that is, the data are not sensitive to atmospheric metallicities  [M/H] > 2.5, with the exception of NH$_3$ for which  the upper limit is already reached at [M/H] > 2.25. For atmospheres obscured by an opacity deck we obtain the strongest constraints for H$_2$O and CH$_4$ where we would be sensitive to signals above opacity decks located at pressures down to $\log_{10}(P_\mathrm{c})\ge -3.5$ for  H$_2$O  and $\log_{10}(P_\mathrm{c})\ge-$5.1 for CH$_4$. This suggests  H$_2$O and CH$_4$ to be good diagnostics for the atmospheric conditions if they are present. For both H$_2$S and NH$_3$ the upper limit for opacity deck pressures is $\log_{10}(P_\mathrm{c})\ge-1.9$ and for CO$_2$ it is $\log_{10}(P_\mathrm{c})\ge-$1.5. For H$_2$S and CO$_2$ our sensitivity for clear atmospheres is also limited towards lower metallicities at [M/H] $< 0.25$ for H$_2$S and [M/H]$< 2$ for CO$_2$, respectively. For CO the injection recovery we are sensitive to the metallicity range 1.5 $\ge$ [M/H] $\le$ 2 and up to pressures of $\log_{10}(P_\mathrm{c})\ge-1.9$.\\
As we obtained a signal of S/N=3.6 for CO$_2$, the injection recovery map of this molecule needs to be interpreted  in a slightly different way. Here, we cannot easily exclude the parameter range at which the injection recovery test suggests we would have been able to obtain a S/N=3 to 4 detection. This would suggest, rather, that the observed signal may possibly arise from an atmosphere compatible with this S/N range in the injection recovery map. We further explore this possibility in Sect. \ref{sec:co2investigation}.\\
\indent A general limitation of the injection recovery test are the equilibrium chemistry assumptions made for the injected models and the use of a fixed C/O ratio. While only one molecule was tested at a time, the mean molecular weight, $\mu$, was calculated for each model assuming the presence of all other molecules in chemical equilibrium. In Fig.~\ref{fig:mu}, we show the change of $\mu$ as a function of metallicity for different assumed C/O ratios. It shows that for C/O at and above solar C/O=0.55, the mean molecular weight is very similar, but its behaviour with increasing metallicity diverges for C/O values below solar. The limits of our injection recovery grids can therefore still be considered as valid for chemical equilibrium with C/O ratios of solar and above but may be slightly inaccurate for lower C/O values.

\subsection{Investigation of the CO$_2$ signal}\label{sec:co2investigation}
We identified a signal in the CO$_2$ cross-correlation analysis with a peak amplitude of S/N=3.6, which lies slightly below our adopted threshold for a robust detection of S/N=4. We found that with our eight transit data set no molecular signals reach this threshold, indicating that further observations will be needed to achieve a robust high-resolution detection of the atmosphere of GJ\,1214\,b in this band pass. Nevertheless, as the data acquired accumulates towards the amount necessary for a confident detection, signals below the threshold such as the one we report here are expected to appear. In the following section, we examined this potential CO$_2$ feature in more detail to assess the probability of it originating from the planet's atmosphere rather than noise. We also aim to explore the implications it would have if it does originate from the planet's atmosphere. While in the following we refer to this signal, emerging from the comparison of the data with models containing CO$_2$, as the 'CO$_2$ signal', we note that its origin cannot yet be robustly established due to its low S/N.

\subsubsection{CO$_2$ signal decomposition}\label{sec:signaldecomposition}
\begin{figure}
\centering
\includegraphics[width=1\hsize]{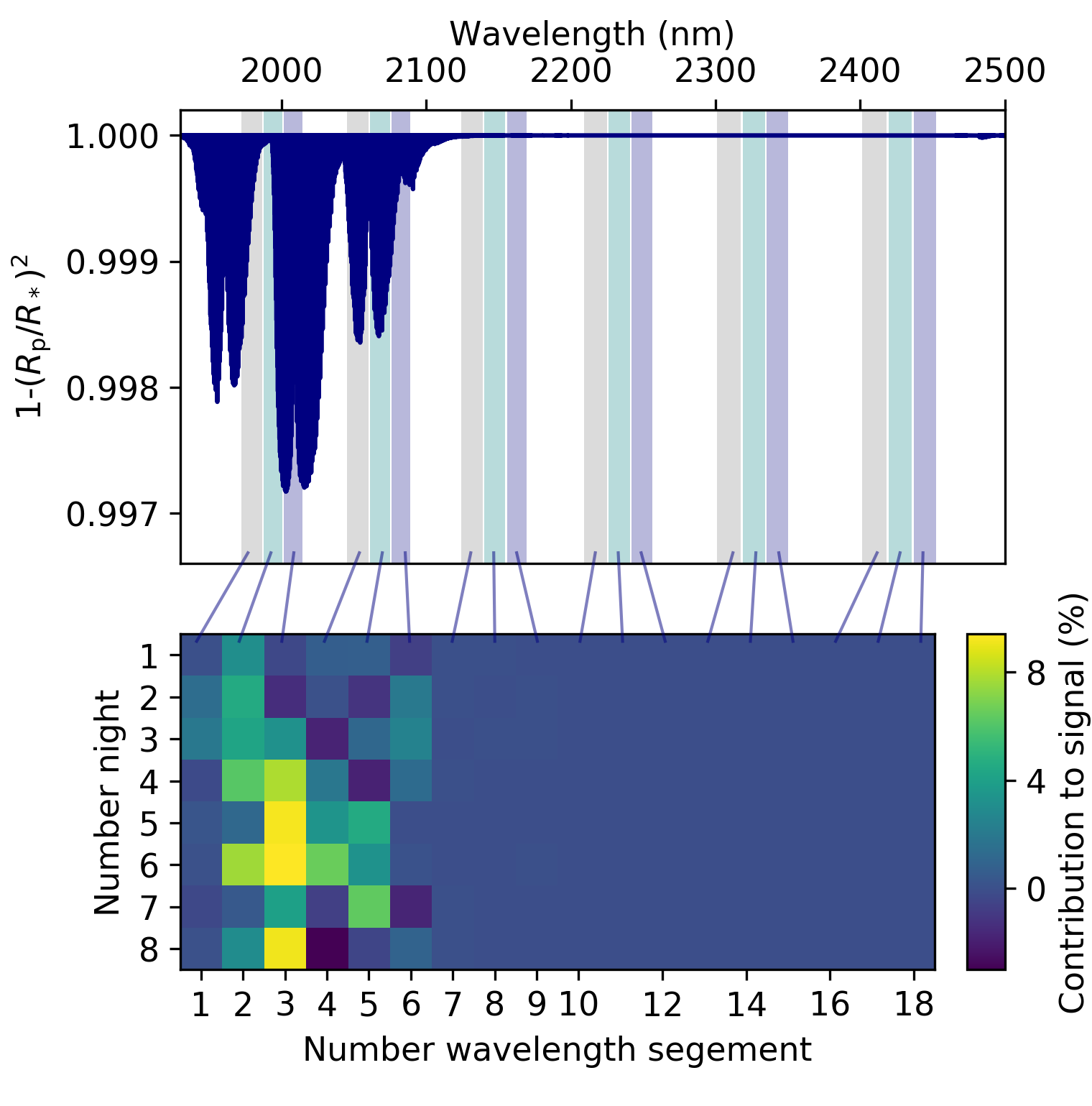}
  \caption{Decomposition of the CO$_2$ signal into its individual contributions. Top panel: CO$_2$ template spectrum (dark blue) and the 18 wavelength segments covered by the K2148 setting (shaded regions). Bottom panel: Decomposition of the S/N=3.6 peak signal of the CO$_2$ CCF into its contributions from individual nights and wavelength segments given in terms of the per cent of the cumulative signal. Wavelength segments in the top panel and corresponding contribution in the bottom panel are indicated with blue lines connecting the two panels for easier identification.} 
     \label{fig:signal_decomposed}
\end{figure}
The CO$_2$ CCF signal is made up of the contributions of eight transit observations with 18 wavelength segments each (only six of which actually cover regions with CO$_2$ lines). We decomposed the signal into its individual components to investigate where it is coming from. For this purpose, we calculated the \mbox{$K_\mathrm{p}$ - $v$} maps for each individual segment of each individual night while normalising each of them by the noise level found in the combined map. We further divided the contribution of each component by the maximum peak signal strength of 3.6 to yield the contributions in units of  per cent of the total signal. The purpose of this exercise was to test if the signal stems only from one night or is caused by any  other similar outlier, which would be indicative of a spurious source of the signal. The result of the decomposition is shown in Fig.~\ref{fig:signal_decomposed}. It shows that several nights contribute to the signal and that the majority of the signal stems from wavelength segment 3, which covers the deepest features of the CO$_2$ band. This is in line with expectations for a real planetary CO$_2$ signal.\\ 
\indent We observed that nights 1 and 2 do not significantly contribute to the overall signal. These are the two nights during which the planetary and telluric lines overlapped during the observations (see Fig.~\ref{fig:obs_stats}). Their lack of contribution to the overall signal excludes the possibility that residual telluric lines are a source of the observed signal. On all other nights, the telluric and planet lines were well separated. The lack of contribution from nights 1 and 2 can be explained by the fact that we mask the deepest telluric lines, that is, the regions that fall below 20\% of the continuum flux. This likely also masked most or all of the planetary lines in these two nights. The masked regions in wavelength segment 3 ranged up to $\pm 20$ km s$^{-1}$ from the centre of the telluric CO$_2$ lines thus fully encompassed the location of the planet signals for nights 1 and 2, at separations of $5.6$ and $3.5$~km s$^{-1}$, respectively. For comparison, on the other nights, the planet's lines were separated by much larger velocity shifts between $-27.1$ to $-45.4$~km s$^{-1}$ (see. Fig.~\ref{fig:obs_stats}). In Fig.~\ref{fig:analysis_alln}, we show the location of the planet CO$_2$ lines with respect to the masked regions in wavelength segment 3 for all eight nights.
Furthermore, any planet lines close to weaker, unmasked CO$_2$ lines would have been more readily removed by \texttt{SYSREM} in these nights due to their proximity to the features being corrected for. The lack of contribution of nights 1 and 2  to the cumulative planetary signal is therefore expected, but it also highlights that the choice of observing night has a major impact on the detectability of signals. In particular the separation between planetary lines and telluric lines required for CO$_2$ studies may be higher than commonly assumed in observational planning. We note that excluding these two nights from the entire analysis (including the analyses done for other molecules) did not impact the overall results obtained within the framework of this study. We further found that nights 3 and 7 did not contribute as much to the signal as other nights. For night 7, this could be explained by the low S/N or the variable humidity of this night. The only unexpected result in this decomposition is the weak contribution from wavelength segment 3 of night 3, despite night 3 having the best observing conditions in our sample, with very low humidity and a S/N of the observed stellar spectra of 100. 
It is important to note, however, that the  $K_\mathrm{p}$ - $v$ maps of individual nights and wavelength segments are still heavily noise dominated. In such maps, the noise at the location of the signal may be high and, by chance, negative, partially cancelling large parts of the true signal. This effect is only mitigated statistically by combining multiple segments and nights. We show the CO$_2$  $K_\mathrm{p}$ - $v$ maps around the expected planet signal for the individual nights in Fig.~\ref{fig:kpmaps_allnights}.

\begin{figure*}
\centering
\includegraphics[width=1\hsize]{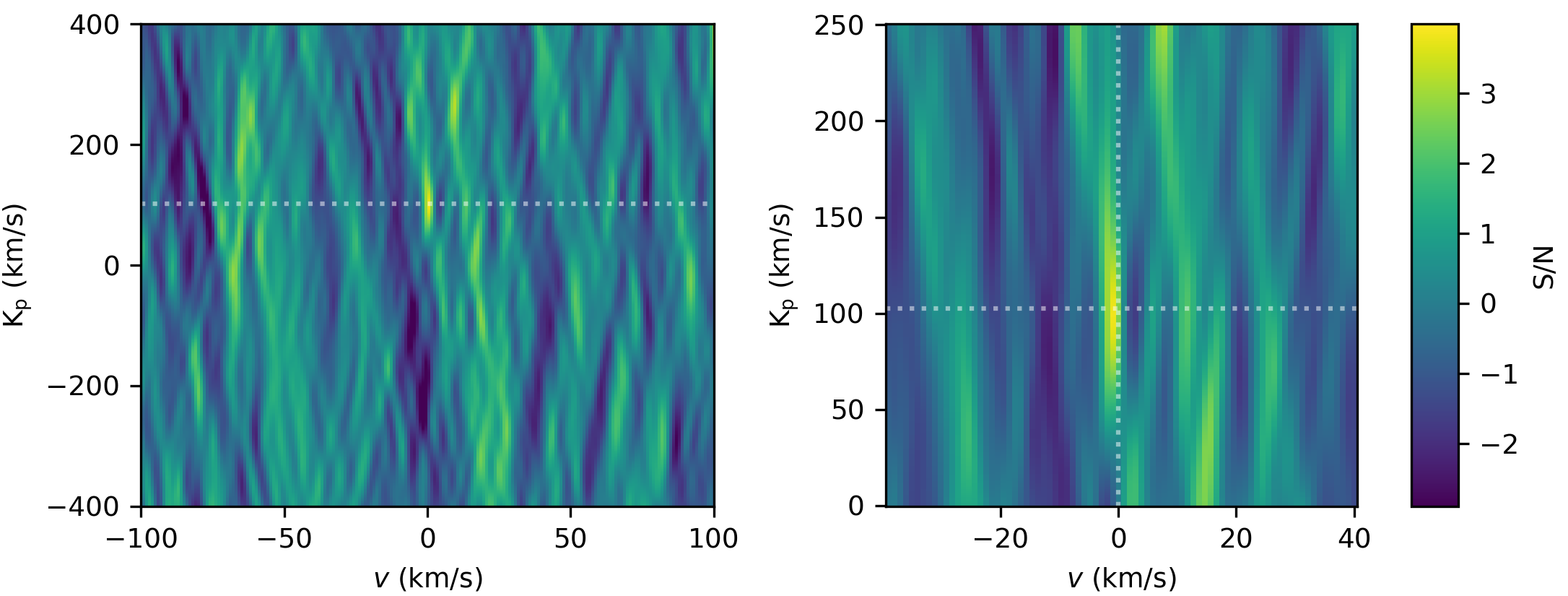}
  \caption{Full $K_\mathrm{p}$ - $v$ map of GJ\,1214\,b. Left panel: calculated with a CO$_2$ model at [M/H]=1.5 and no clouds normalised and shown with extending the normalisation range to the negative $K_\mathrm{p}$ range, and excluding nights 1 and 2. Right panel: Same as left but zoomed in to the peak CO$_2$ signal. The dotted white lines indicate literature $K_\mathrm{p}$ (horizontal) and $v=0$ (vertical).}
     \label{fig:negKp6n}
\end{figure*}

\subsubsection{Model template and $K_\mathrm{p}$ - $v$ normalisation range dependence}
The exact value of S/N=3.6 we achieve for the CO$_2$ signal is subject to choices made during the analysis, such as the $K_\mathrm{p}$ and $v$ range explored and the regions chosen for normalisation. For example, the peak  S/N  value increased to 3.7 when we expanded our exploration and consequently the normalisation range to the negative $K_\mathrm{p}$ space, from $-400$ to $+400$ km s$^{-1}$ as suggested by \citet{SanchezLopez2019,SanchezLopez2020}. It further increased to 3.8 (see Fig.~\ref{fig:negKp6n}) when we omitted nights 1 and 2, a measure that can be argued for considering that these nights cannot significantly contribute to a planetary CO$_2$ signal due to overlap with wide masked telluric regions as discussed in Sect.\ref{sec:signaldecomposition}. A poor choice of the synthetic template used for the cross-correlation on the other hand, may decrease the signal amplitude.  Our exploration of the effect of different template spectra on the CCF results showed that the signal amplitude of the CO$_2$ signal in the \mbox{$K_\mathrm{p}$ - $v$} maps varied with the template model used. Fig.~\ref{fig:detectiongrid} shows that we obtained S/N $\ge$ 3.56  for models with metallicities  [M/H] < 1.75 but that the CCF signal amplitude rapidly drops off for model templates with higher metallicities. Contrarily, the CCF results for CO$_2$ appear to not be very sensitive to the presence of an opacity deck, with peak S/N values being achieved for metallicities of [M/H] = 1.5 regardless of the height of the opacity deck. While this suggests that the line positions and ratios of [M/H] = 1.5 models match the data best, it is not a good indicator for identifying the atmosphere that best explains the data, since the CCF analysis is not sensitive to the overall line amplitudes. Moreover, the tested grid of models is limited in its fixed C/O ratio and $T$-$p$ profile as well as its discretely sampled metallicities and opacity deck pressures. A much better way to determine which model best explains the observed signal is to run a Bayesian retrieval on the data.

\begin{table}[ht]\renewcommand{\arraystretch}{1.5}
\centering
 \caption[]{Results of the atmospheric retrievals.}\label{tab:rerivalresults}
\begin{tabular}{llll}
 \hline \hline
  Parameter &
  Prior range &
  Median value &
  Unit
 \\ \hline
\multicolumn{3}{l}{\textit{Eq. chem., CO$_2$-only retrieval}} &\\
$T_\mathrm{iso}$           & [200, 800]    & $509.23 ^{+102.0}_{-58.68}$     &  K    \\ 
$\log_{10}(P_\mathrm{c})$ & [$-7.9$, 2]     & $-0.88 ^{+1.95}_{-2.48}$       &  $\log_{10}$(bar)  \\
C/O                       & [0, 1.5]      & $0.63 ^{+0.54}_{-0.41}$          &  ... \\
$[$M/H$]$                & [$-1$, 4]       & $1.51 ^{+0.68}_{-0.75}$        & ...  \\
$K_\mathrm{p}$                     & [50, 150]   & $107.04 ^{+22.88}_{-19.38}$          & km\,s$^{-1}$  \\
$v_0$                     & [$-3.5$, 0.5]   & $-1.46 ^{+0.55}_{-0.64}$         & km\,s$^{-1}$  \\
$\beta$                   & [0.1, 2]      & $1.0251(8)$     & ...   \\
 \textit{} & &   & \\
 \multicolumn{3}{l}{\textit{Free chem.,   retrieval of seven molecules}} &\\
 $T_\mathrm{iso}$           & [100, 1000]    & $397.63 ^{+282.62}_{-197.11}$    &  K    \\
$\log_{10}(P_\mathrm{c})$ & [$-7.9$, 2]       &  $-3.04 ^{+2.52}_{-1.53}$       &  $\log_{10}$(bar)  \\ 
$\log_{10} \mathrm{VMR}_\mathrm{H2O}$ 
& [$-7.9$, 2]     &  $-4.68 ^{+2.34}_{-9.64}$      &  .. \\
$\log_{10}(\mathrm{VMR_{CH_4}})$ & [$-20$, $-0.1$]      &  $-11.7 ^{+5.33}_{-5.46}$       &  .. \\
$\log_{10}(\mathrm{VMR_{CO}})$ & [$-20$, $-0.1$]      &  $-10.92 ^{+7.24}_{-6.25}$        &  .. \\
$\log_{10}(\mathrm{VMR_{CO_2}})$ & [$-20$, $-0.1$]     &  $-3.2 ^{+1.12}_{-2.26}$        &  .. \\
$\log_{10}(\mathrm{VMR_{H_2S}})$ & [$-20$, $-0.1$]     &  $-12.37 ^{+5.46}_{-5.23}$       &  .. \\
$\log_{10}(\mathrm{VMR_{NH_3}})$ & [$-20$, $-0.1$]      &  $-11.32 ^{+5.2}_{-5.94}$       &  .. \\
$\log_{10}(\mathrm{VMR_{N_2}})$* & [$-20$, $-0.1$]      &  $-10.93 ^{+6.54}_{-6.2}$        &  .. \\
$\beta$  & [0.1, 2]      & $1.064(5)$     & ..   \\
$\mu$ & calculated      & $2.43 ^{+0.86}_{-0.09}$   &  ..   \\
$\mathrm{[M/H]}$  & calculated      &$0.48 ^{+0.89}_{-1.70}$   & ..   \\
\hline
\end{tabular}
\tablefoot{The uncertainties quoted with the median value correspond to the 16th and 84th percentiles. (*) The VMR of N$_2$ should be interpreted as the combined VMR of all other metals not considered in the retrieval.}
\end{table}

\begin{figure}
\centering
\includegraphics[width=1\hsize]{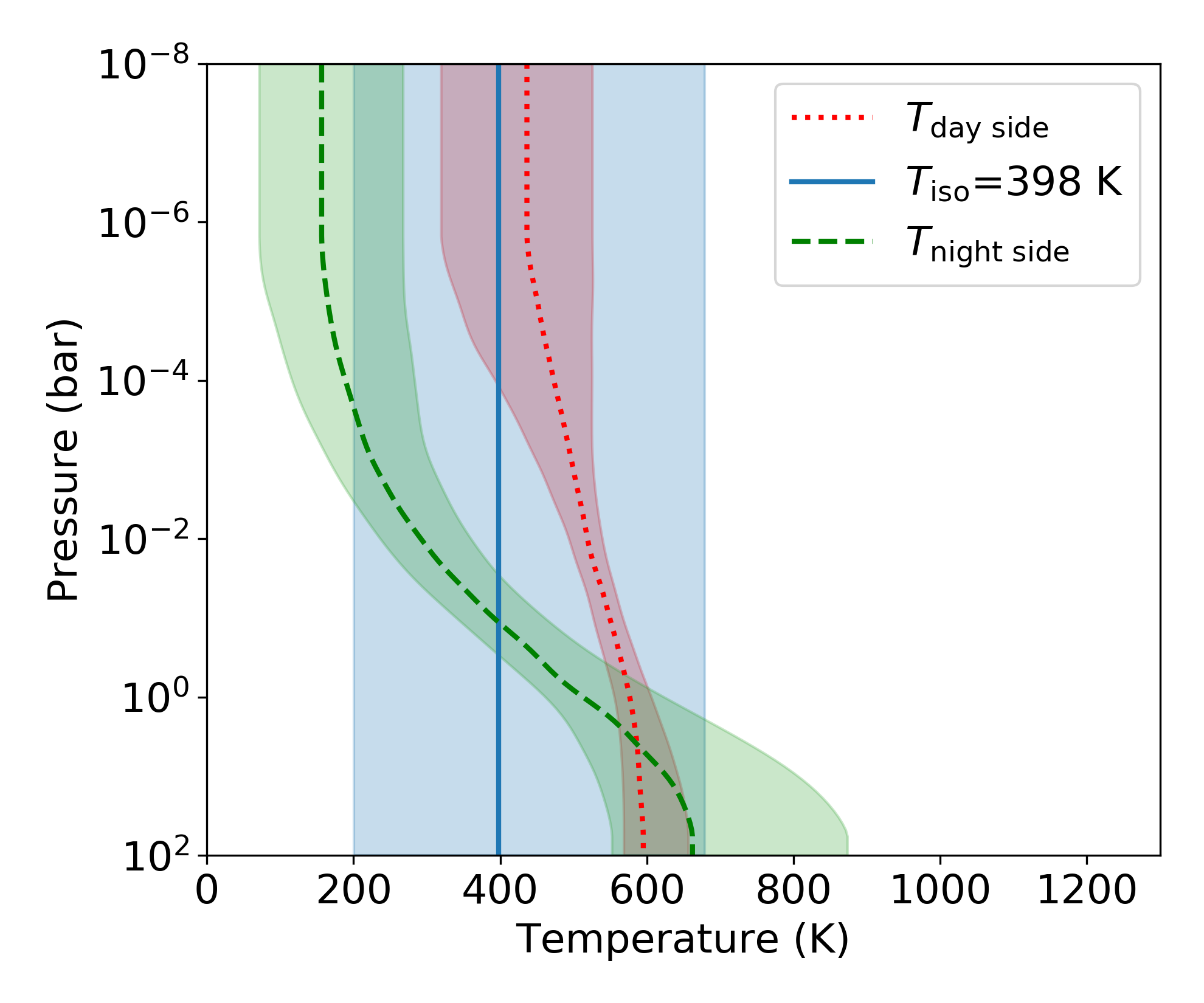}
  \caption{Isothermal temperature-pressure profile retrieved in the free chemistry retrieval. Its $1\sigma$ uncertainty intervals are indicated in blue shaded regions. The profile is compared to the  profiles and their $1\sigma$ uncertainties retrieved by \citet{Kempton2023} from JWST MIRI data of day- and night-side emission. The latter were extracted from their Fig. 6  using \texttt{plotdigitizer} (\url{https://plotdigitizer.com}) and extrapolated for pressures below 10$^{-6}$ bar.}
     \label{fig:retrieved_tp}
\end{figure}
\begin{figure*}
\centering
\includegraphics[width=1\hsize]{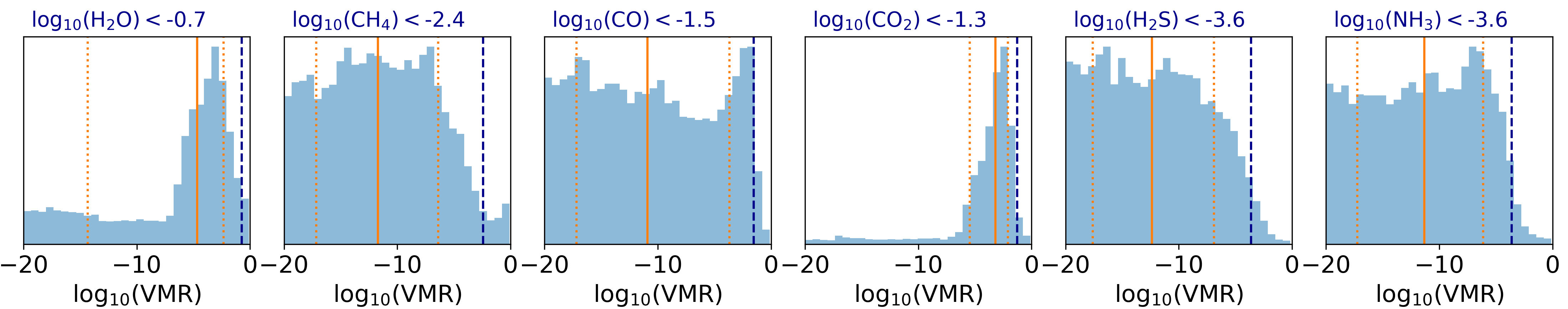}
  \caption{Upper limits for six molecules from the free chemistry retrieval. We show the posterior distributions and mark in dark blue the 2$\sigma$ upper limits (calculated as the 97.5th percentile). The corresponding value is given above the histograms. In orange we mark the median and 1$\sigma$ uncertainty intervals, also listed in Table \ref{tab:rerivalresults}.}
     \label{fig:histro_vmr}
\end{figure*}
\begin{figure*}
\centering
\includegraphics[width=1\hsize]{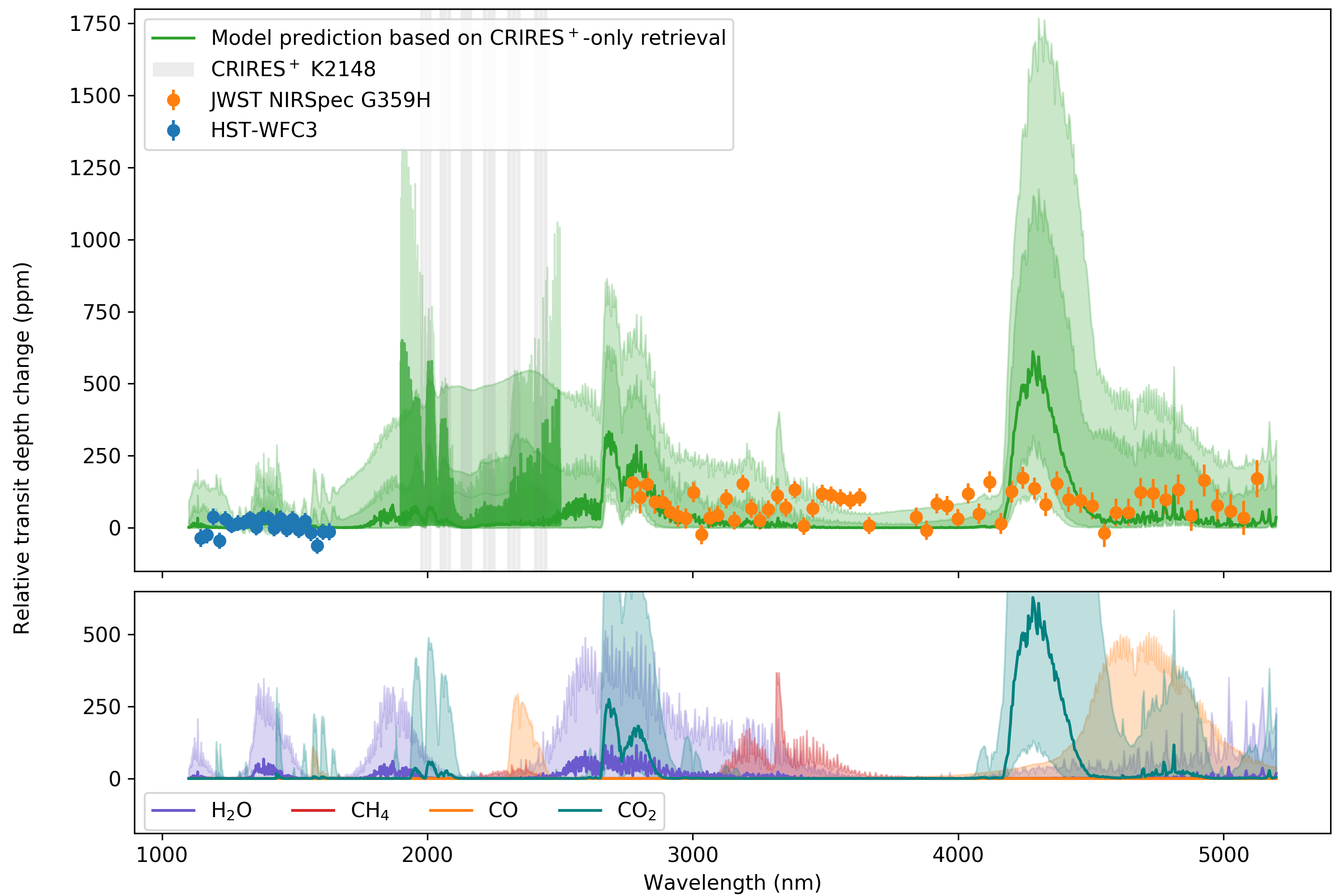}
  \caption{Comparison of the multi-molecule, free-chemistry retrieval models to low-resolution data. Top panel: Predicted relative transit depth change from the free chemistry models retrieved from CRIRES$^+$ data  (green). The model predictions are plotted in high-resolution in the wavelength region overlapping with the CRIRES$^+$ remeasurements indicated by the grey shaded areas and in low-resolution for the rest of the wavelength range. Space-based measurements from HST (blue) and JWST (orange) are also shown. The 1$\sigma$ and 1.5$\sigma$ uncertainties of the model are indicated as darker and lighter green shaded areas, respectively. The HST-WFC3 and JWST NIRSpec G359H measurements were taken from the online data provided by \citet{Schlawin2024} containing the HST data adapted from \cite{Kreidberg2014}. Bottom panel: The contributions to the model by the opacities of the individual molecules are shown including the 1$\sigma$ and 1.5$\sigma$ uncertainties. H$_2$S and NH$_3$ are not shown as they do not contribute to the opacities of the global model in any of the shown wavelength regions based on the retrieved VMRs.
  }
     \label{fig:retmodel_vs_jwst_freechem} 
\end{figure*}

\subsection{Results of the Bayesian retrieval}\label{sec:retrieval_results_discussion}
We applied Bayesian retrievals to study the potential CO$_2$ signal and place upper limits on the presence of other species. As a first step, we performed a simplified retrieval assuming CO$_2$ and CIA absorption as the only opacity sources under equilibrium chemistry. We then expanded our modelling to a multi-molecule retrieval including the opacities of all six molecules investigated in the cross-correlation study, with a free chemistry set-up. 
\subsubsection{Results of the equilibrium chemistry CO$_2$-only retrieval}
The retrieved orbital parameters \mbox{$K_\mathrm{p}=107^{+23}_{-19}$ km s$^{-1}$} and \mbox{$v_0=-1.46^{+0.55}_{-0.64}$ km s$^{-1}$} agree with the peak position found in the CCF analysis and with the values expected for GJ\,1214\,b based on literature parameters.  
We retrieved an isothermal temperature of $T_\mathrm{iso}=509^{+102}_{-59}$ K, which lies slightly above the day-side upper-atmosphere temperatures of $T$-$p$ profiles retrieved from JWST emission spectroscopy data.  We further retrieved a metallicity of [M/H]=$1.51^{+0.68}_{-0.75}$, and a relatively higher pressure of the opacity deck at $\log_{10}(P_\mathrm{c})=-0.88^{+1.95}_{-2.48}$. 
The \mbox{CO$_2$-only} set-up failed to constrain the C/O ratio (for which we got \mbox{C/O= $0.63^{+0.54}_{-0.41}$}), which is to be expected when no molecular abundance ratios are evaluated. 
The offset of $v_0$ from zero with respect to the expected planet rest frame found for the signal could be indicative of atmospheric dynamics. In the case of a blue shift, as seen here, for a day- to night-side flow. This scenario would be compatible with the measured temperature corresponding closer to  GJ\,1214\,b's day-side rather than its night-side temperature. The shift may however also arise from small uncertainties in the planet ephemeris \citep{Meziani2025}. 
In the top right panel of Fig.~\ref{fig:retrivalresults} we show the contours of the probability distributions of the retrieved metallicity and cloud deck pressure overplotted on the S/N grid for CO$_2$ from Fig.~\ref{fig:snrgrid}, which shows the sensitivity of our data to specific models. The direct comparison shows that the retrieved distribution overlaps with the regions at which injection-recovery tests expect us to achieve a signal detection of \mbox{S/N= 3 - 4}. This demonstrates consistency between our observed S/N = 3.6 signal and the sensitivity of our data to the atmospheric conditions most compatible with the signal according to the retrieval.\\
\indent The retrieved atmospheric scenario is not immediately reconciled with the absence of other molecular detections by the CCF analysis. Based on the injection-recovery tests, and assuming equilibrium chemistry, the retrieved $1\sigma$ ranges of metallicity and opacity deck pressure would imply the detectability at S/N$\ge 4$ for all other tested molecules. However,  the top right panel of Fig.~\ref{fig:retrivalresults} shows that within the $2\sigma$ uncertainties of the retrieval there exist high-metallicity (up to [M/H] =3) and high-altitude opacity deck scenarios ($\log_{10}(P_\mathrm{c})=-5.5$) which would be consistent with the non-detections of other molecules. Nonetheless, the absence of clear signals of other molecules in the CCF analysis likely indicates that to capture the true atmosphere of GJ\,1214\,b disequilibrium chemistry and non-grey clouds and hazes will have to be considered.
\subsubsection{Results of the free chemistry multi-molecule retrieval}
To investigate atmospheric scenarios that allow for the CO$_2$ signal and non-detections of other molecules to co-exist, we employed a free-chemistry retrieval including all molecules tested for in Sec.~\ref{sec:ccf}, plus N$_2$ as a proxy for species that are not exhibiting any absorption in the investigated wavelength range but contribute to the mean molecular weight. The freedom of this retrieval can capture some of the effects of disequilibrium chemistry, which can deplete or enrich the abundances in the upper atmosphere, for example through photochemistry.  We retrieved an isothermal temperature of  $T_\mathrm{iso}= 398 ^{+283}_{-197}$, slightly lower than the one obtained in the equilibrium chemistry retrieval and with larger uncertainties. This highlights that when allowing more freedom in the molecular abundances a larger range of temperatures can explain the results. The resulting isothermal \mbox{$T$-$p$ profile} and its $1\sigma$ uncertainty intervals are shown in Fig.~\ref{fig:retrieved_tp}, together with the temperature-pressure profiles for the planet’s day and night sides retrieved from JWST observations by \citet{Kempton2023}. The retrieved temperature lies between the temperatures of these profiles in the low-pressure regime, to which our observations are most sensitive. This follows expectations for the temperature of the  terminator at the transition between the day and night sides. Contrary to the CO$_2$-only retrieval, the multi-molecule retrieval converges to much lower pressures for a potential cloud or haze deck ($\log_{10}(P_\mathrm{c})=-3.04^{+2.53}_{-1.53}$), which is more aligned with results of previous studies that found high haze layers to be the most likely explanation for the muted spectral features \citep{Kreidberg2014, Gao2023, Schlawin2024, Ohno2025}. This can be explained by the fact that the CO$_2$-only retrieval neglects the opacities of other molecular species and therefore does not require high-altitude clouds or hazes to obscure their spectral features. The free-chemistry retrieval, which includes these opacities, allows us to constrain the VMRs of all tested molecules. 
These can be interpreted as the abundances at higher altitudes, which we are sensitive to, rather than global abundances or those found in deeper layers of the atmosphere. For CO$_2$ we obtain a well constrained VMR of \mbox{$\log_{10} (\mathrm{VMR_{CO_2}})=-3.2 ^{+1.12}_{-2.26}$}. For the other molecules the retrieval can only place relatively weak constraints, with the VMRs spanning a wide range from very low to relatively high values (see Figs.~\ref{fig:retrivalresults_freechem} and \ref{fig:histro_vmr}). In the case of CO the highest abundances appear to correlate with very low temperatures around $T_\mathrm{iso}=100$ K, outside of the temperature range preferred by the retrieval. The retrieval finds stronger than expected constraints on the VMR of H$_2$O. While very low VMRs are still possible within its 1$\sigma$ uncertainties the retrieval shows a clear preference for intermediate values of $\log_{10}(\mathrm{VMR_{H_2O}})=-4.68^{+2.34}_{-9.64}$. This is in alignment with the observed S/N of 1.49 in the $K_\mathrm{p}$-$v$ map of H$_2$O at the location of the CO$_2$ signal (see Table \ref{tab:CCFpeaks}).  Given that this CCF signal does not present as a clearly isolated peak in the map and that the lower range for $\log_{10}(\mathrm{VMR_{H_2O}})$ is still sufficiently represented in the posterior distribution, we consider this value still an upper limit for water, rather than evidence for its detection. We derived $2\sigma$ upper limits for all the tested molecules from their posterior distributions, which are indicated in Fig.~\ref{fig:histro_vmr}.\\
\indent Using the retrieved VMRs we calculated the probability distributions of the mean molecular weight and the metallicity of the atmosphere. This revealed that the combinations of VMRs preferred by the retrieval correspond to relatively low metallicities ([M/H]$= 0.48^{+0.89}_{-1.70}$) and, consequently, low mean molecular weights ($\mu= 2.43^{+0.86}_{-0.09}$). It therefore appears that the retrieval explains the lack of other strong molecular features via a reduced temperature and increased height of the opacity deck rather than forcing either extremely low VMRs (absence of the species) or very large VMRs (high metallicities and thus low scale heights). 
These retrieved metallicities stand in contrast to the results from retrievals based on low-resolution observations, who propose ultra-high metallicities (e.g. [M/H]=3.69, \citeauthor{Ohno2025} \citeyear{Ohno2025}). In such studies models with low mean molecular weights are often disfavoured due to the lack of strong broad band haze features that should be visible in these scenarios. Therefore, a likely explanation for this difference is that our high-resolution data is not sensitive to such features and therefore does not penalise low mean molecular weight models for their absence. The relatively large uncertainties towards higher metallicities obtained by the retrieval, which coincide with the highest VMRs values sampled for CO$_2$, suggest that moderately metal rich scenarios are still compatible with the results, if the CO$_2$ abundance is high.
 \subsubsection{General remarks on the retrieval results}
Generally, the retrieved values for the temperatures and the atmospheric conditions found in the retrievals correspond to conditions expected for sub-Neptune planets \citep{Lavvas2019, Davenport2025}.  The retrieved scaling parameter for the data uncertainties converged to $\beta=1.0251(8)$, for the CO$_2$ only and to $\beta=1.064(5)$ for the multi-molecule  retrieval indicating that the uncertainties provided by the \texttt{CR2RES} pipeline version 1.4.2 are accurate and our derived uncertainty estimates are reasonable. The results and the prior ranges of all parameters are summarised in Table \ref{tab:rerivalresults} and the probability distributions are shown in Fig.~\ref{fig:retrivalresults} and Fig.\ref{fig:retrivalresults_freechem}. \\
\indent The moderately enriched metallicities preferred by the retrievals suggest a different atmospheric scenario from those proposed in previous studies, where the muted spectral features obtained a low-resolution were attributed to not only high-altitude aerosols but also high metallicities \citep{Kreidberg2014, Gao2023, Schlawin2024, Ohno2025}. We compare our retrieved model to the space-based low-resolution data obtained with  HST-WFC3  and JWST NIRSpec G359H in Sect. \ref{sec:comparetojwst}.

\begin{figure*}
\centering
\includegraphics[width=1\hsize]{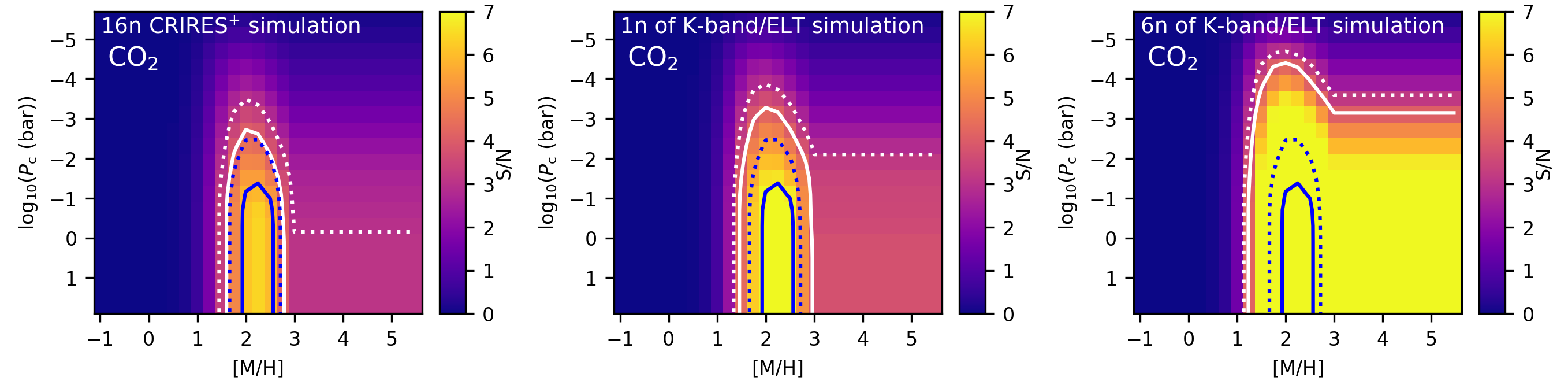}
  \caption{
  Simulated S/N grids showing the expected constraints that can be obtained by obtaining more data. The dotted white lines represent the contours of the simulated data, with S/N=3 (dotted) and S/N=4 (solid) respectively. The dark blue lines represent the contours of the injection recovery tests of our real eight transit data set shown in Fig \ref{fig:snrgrid}. The constraints are shown for 16 CRIRES$^+$ transits, that is, another eight transits with CRIRES$^+$ added to the eight nights presented in this manuscript (left panel), for one transit and six transits observed with a fictitious K-band spectrograph at the ESO-ELT (middle panel and right panel, respectively.} 
     \label{fig:futuresim}
\end{figure*}

\subsubsection{Comparison of CO$_2$ signal amplitude with HST and JWST data}\label{sec:comparetojwst}
While our models in the Bayesian retrievals are of limited complexity, they give us an idea of the amplitude of the CO$_2$ feature in the K band that is compatible with our measured signal and place upper limits on the amplitudes of features from other molecules. 
This allows us to predict the corresponding amplitudes of the features of CO$_2$ and the other tested molecules in the L and M bands ($\sim$ 2.7 to 5.3 $\mathrm{\mu}$m) and compare them to space based low-resolution observations.\\
\indent In Figs.~\ref{fig:retmodel_vs_jwst_freechem} and ~\ref{fig:retmodel_vs_jwst}, we show the relative transit depth changes predicted by the two retrieved models and their uncertainty intervals extrapolated to a larger wavelength range (1000-5200 nm) at lower spectral resolution together with the  data of HST-WFC3  and JWST NIRSpec G359H data obtained by \citet{Kreidberg2014} and \citet{Schlawin2024}, respectively. We used the JWST NIRSpec data reduced using \texttt{tshirt!} and the version of the HST data provided by \citet[][see their Fig. 9]{Schlawin2024} who derived a downward correction for the HST data by 119 ppm from broadband light curve fitting with new orbital parameters. We note that we did not apply the additional 56 ppm upwards shift those authors applied to the HST data afterwards, motivated by their own retrieval.\\
\indent It is evident that the amplitudes of the CO$_2$ features  at 2.8 $\mathrm{\mu m}$ and 4.3 $\mathrm{\mu m}$ in the JWST data are lower than those predicted by our best models but they are compatible within our models' $1.5\sigma$ uncertainties.  
An explanation for the difference in signal amplitudes between the K- and M-band CO$_2$ features from the predictions could be significant deviations of the \mbox{$T$-$p$ profile} from the isothermal approximation as well as  non-equilibrium chemistry causing deviations in the  altitude-dependent volume mixing ratios via photodissociation. This would lead to different abundance profiles than those we calculated with \texttt{easyCHEM} for the equilibrium chemistry retrieval but also those of the free chemistry retrieval where we kept abundances constant with altitude. Finally, when considering low-resolution spectroscopy and large wavelength ranges the assumption of grey clouds and hazes is insufficient. A wavelength-dependent continuum absorption by clouds or hazes may explain a difference in relative CO$_2$ absorption depth between the different band passes. We note that even the amplitude ratio between the potential CO$_2$ features at 2.8 and 4.3~$\mathrm{\mu m}$ in the JWST data does not necessarily match model expectations, with the 2.8~$\mathrm{\mu m}$ feature appearing stronger relative to the 4.3~$\mathrm{\mu m}$ feature than predicted \citep{Ohno2025}. One possible source of non-grey high-altitude opacity that is showing different levels of absorption between the 1.9 $\mu$m and 3-5 $\mu$m range are tholin hazes, which have been proposed as a contributor to featureless spectra of sub-Neptunes \citep{Khare1984,Adams2019,Kawashima2019b,He2024}. However, accounting for the wavelength dependence of realistic hazes and clouds introduces its own challenges to the interpretation of featureless spectra at low-resolution. In lower-metallicity scenarios their broadband features can become too large to remain compatible with the flat spectra observed. This is one of the factors that have driven the preference for high metallicities in the previous interpretations of GJ\,1214\,b's low-resolution spectrum \citep{Ohno2025}. Intense effort has been devoted into the modelling of the low-resolution observations \citep[e.g.][]{MRKempton2012,Morley2013,GaoBenneke2018,HuSeager2014,Charnay2015,Morley2015,OhnoOkuzumi2018,Adams2019,Kawashima2019a,Lavvas2019,Ohno2020,Christie2022,Gao2023,Nixon2024,Ohno2025,Malsky2025,Li2025}. Extending such modelling to simultaneously incorporate both the high-resolution and the low-resolution data at the same level of complexity lies beyond the scope of this work. Nevertheless, such a study would undoubtedly benefit from improved detection significances through additional data on both the high-resolution and low-resolution sides, given that even the JWST detection of CO$_2$ has been reported at a marginal significance of 2.4$\sigma$ \citep{Schlawin2024}. In the following (Sect. \ref{sec:futuresim}) we investigate how future high-resolution observations could improve our constraints on GJ\,1214\,b's atmosphere.

\subsection{Results of simulated future observations}\label{sec:futuresim}
We simulated the improvements on the detectability of features that could be achieved by adding eight more CRIRES$^{+}$ observing nights. The results are shown in Fig.~\ref{fig:futuresim} and show that with eight more CRIRES$^+$ transits the models previously reaching \mbox{S/N = 3} now reach \mbox{S/N = 4}. 
If the CO$_2$ signal measured in this work stems from the planet atmosphere additional nights should consequently allow it to be confirmed. 
We also see that the additional data would raise the sensitivity towards atmospheric scenarios with higher metallicities as they allow for sub-threshold S/N = 3 detections of signals up to at least [M/H] = 5.5 for clear atmospheres. Given that two of the original transits are unsuitable for CO$_2$ studies as discussed in Sect.~\ref{sec:signaldecomposition}, the results for this molecule may, in reality, reflect the detection significance obtainable with 12 rather than 16 transits, with minor added noise due to the four unsuitable nights.\\ 
\indent The simulations of a hypothetical K-band spectrograph at the ELT show that with one ELT transit we could achieve a similar improvement to the one going from 8 to 16 CRIRES$^+$  nights, that is, signals detectable at S/N=3 with 16 CRIRES$^+$ nights would be detectable at S/N=4 for metallicities [M/H]$\le 3$. 
Above [M/H]=3, even an ELT transit observation is not sufficient to reach the S/N=4 threshold, and the S/N=3 threshold only shifts slightly to also encompass moderately high cloud decks. However, expanding our hypothetical observations to six nights at the ELT finally facilitates detections of S/N$\ge 4$ for atmospheres with extremely high metallicities (up to at least [M/H]=5.5) even for cloud deck pressures as low as $\log_{10}(P_\mathrm{c})=-3$.

\section{Summary and conclusions}\label{sec:summaryconclusons}
We studied the elusive transmission spectrum of the canonical sub-Neptune GJ\,1214\,b using eight transit observations obtained at high-resolution with CRIRES$^+$. Our results are listed below:
\begin{itemize}
    \item We obtained non-detections of H$_2$O, CH$_4$, CO, H$_2$S, and NH$_3$ for a range of model templates. 
    \item Injection recovery tests demonstrated that the data are not sensitive to metallicities above [M/H]$=2.5$ and opacity deck pressures lower than $\log_{10}(P_\mathrm{c}) =$ -5.1.
    \item We found a S/N=3.6 CCF signal for the analysis with a CO$_2$ template spectrum, moving in time with the orbital velocity of GJ\,1214\,b at a small blue-shifted velocity offset:\ \mbox{$v_0=-1$ km s$^{-1}$} based on the CCF analysis; \mbox{$v_0=-1.46^{+0.55}_{-0.64}$ km s$^{-1}$} based on the retrieval.
    
    \item A Bayesian retrieval of this CO$_2$ signal with a single-species (equilibrium chemistry) set-up resulted in an isothermal temperature of  
    $T_{\mathrm{iso}} = 509^{+102}_{-59}$ K, slightly above the day-side upper-atmosphere temperature from $T$-$p$ profiles retrieved from JWST emission spectroscopy data at low pressures, a metallicity of [M/H]$ = 1.51^{+0.68}_{-0.75}$, and an opacity deck at $\log_{10}(P_\mathrm{c}) = -0.7^{+1.9}_{-2.5}$. 
    \item A Bayesian retrieval including opacities of multiple molecular species (free-chemistry scenario) resulted in an isothermal temperature of 
    $T_{\mathrm{iso}} = 397^{+183}_{-197}$ K, a metallicity of [M/H]$ = 0.48^{+0.89}_{-1.70}$ derived from the retrieved VMRs, and an opacity deck at $\log_{10}(P_\mathrm{c}) = -3.04^{+2.52}_{-1.53}$. The retrieved abundance for CO$_2$ is $\log_{10}(\mathrm{VMR_{CO_2}}) = -3.2^{+1.12}_{-2.2}$; whereas upper limits were obtained for the other molecules (H$_2$O, CH$_4$, CO, H$_2$S, and NH$_3$). The retrieved temperature is compatible with a value intermediate between the upper-atmosphere temperatures from JWST-derived day- and night-side $T$-$p$ profiles, as would be expected for the planetary terminator.   
    \item Predictions for the amplitude of the CO$_2$ features at 2.8$\mathrm{\mu}$m and $4.3\mathrm{\mu}$m based on our simplified retrieved models are greater than what has been derived from observations; however, these predictions are  still compatible with the JWST NIRSpec G359H observations within the models'  $1.5\sigma$ uncertainties.
\end{itemize}
Under simplified equilibrium chemistry modelling assumptions, the amplitude of the measured CO$_2$ signal implies opacity deck pressures and metallicities, for which additional molecular features are also expected to be detectable. Under free chemistry assumptions, the non-detection of other molecules alongside the CO$_2$ detection can be explained with a high-altitude opacity deck at simultaneously high CO$_2$ VMR, without the need for an extremely high atmospheric metallicity. For both scenarios, however, we would expect stronger CO$_2$ signatures in the M band than those detected with JWST. This suggests that more sophisticated modelling of GJ\,1214\,b’s atmosphere is required to reconcile the results from high- and low-resolution observations and to place them into the broader atmospheric context. Moreover, additional observations of the planet should be obtained to confirm the atmospheric signals for both high- and low-resolution studies and to improve the constraints on the retrieved atmospheric parameters.\\
\indent Overall, our study highlights that the challenges of characterising sub-Neptune exoplanets (for which clouds and hazes and higher metallicities mute spectral features) are also relevant with respect to high-resolution studies. Nonetheless, high-resolution data appear to provide a complementary perspective on the atmosphere that is not immediately consistent with the picture painted by low-resolution results, indicating that efforts at combined interpretations will have to be undertaken in the future.

\section{Data availability}
The extracted CRIRES$^+$ spectra, including the refined wavelength solution derived with \texttt{MOLECFIT}, are available at \url{https://zenodo.org/records/19387252} \citep{lavail_2026_19387252}. The raw data are avaiable in the ESO archive\footnote{\url{https://archive.eso.org/eso/eso_archive_main.html}} under programme IDs 108.22CH,
108.22PH, 109.23HN, 111.254J and 113.26G.

\begin{acknowledgements}
CRIRES$^+$ is an ESO upgrade project carried out by Thüringer Landessternwarte Tautenburg, Georg-August Universität Göttingen, and Uppsala University. The project is funded by the Federal Ministry of Education and Research (Germany) through Grants 05A11MG3, 05A14MG4, 05A17MG2 and the Knut and Alice Wallenberg Foundation. This project is based on observations collected at the European Organisation for Astronomical Research in the Southern Hemisphere under the ESO programmes 108.22CH, 108.22PH, 109.23HN, 111.254J, 113.26GE. 
D.C. is supported by the LMU-Munich Fraunhofer-Schwarzschild Fellowship and by the Deutsche Forschungsgemeinschaft (DFG, German Research Foundation) under Germany´s Excellence Strategy – EXC 2094 – 390783311. 
F.L. is supported by the European Union (ERC-CoG, EVAPORATOR, Grant agreement No. 101170037). Views and opinions expressed are however those of the author(s) only and do not necessarily reflect those of the European Union or the European Research Council. Neither the European Union nor the granting authority can be held responsible for them.
E.N. acknowledges the support from the Deutsches
Zentrum für Luft- und Raumfahrt (DLR, German Aerospace Center) - project
number 50OP2502. 
A.D.R., L.B.-Ch., and N.P. acknowledge support by the Knut and Alice Wallenberg Foundation (grant 2018.0192). 
OK acknowledges support by the Swedish Research Council (grant agreements no. 2023-03667) and by the Swedish National Space Agency.
M.R. acknowledges the support by the DFG priority program SPP 1992 “Exploring the Diversity of Extrasolar Planets” (DFG PR 36 24602/41 and CZ 222/5-1, respectively).
D.S. acknowledges funding from project PID2021-126365NB-C21(MCI/AEI/FEDER, UE) and financial support from the grant CEX2021-001131-S funded by MCIN/AEI/ 10.13039/501100011033.

\end{acknowledgements}

%
%
\bibliographystyle{aa} 
\bibliography{bib_gj1214crires.bib}

\appendix
\section{Additional material}
\subsection{Data reduction}\label{sec:preproc} 
The spectra were extracted using the CRIRES$^+$ data reduction software \texttt{CR2RES} (Version 1.4.2). The time series observations were sorted into pairs of consecutive A and B spectra, and the pipeline was run on each pair. The reduction steps of the pipeline included dark and flat field correction as well as removal of bad pixels. The sky background was corrected by subtracting the A and B position frames from each other. 
The pipeline also produced a wavelength solution based on wavelength calibration frames taken with a uranium-neon lamp and a Fabry-Perot etalon. We further refined this wavelength solution using the ESO tool \texttt{molecfit} \citep{Smette2015, Kausch2015}. \texttt{Molecfit} can fit telluric lines in high-resolution spectra and provide an improved wavelength solution that is optimised as part of the fitting procedure. We used it to obtain an improved wavelength solution for each wavelength segment of the CRIRES$^+$ spectra for one reference spectrum in each of the two nodding positions for every night. We found that for night 3 the full width half maximum (FWHM) of the point spread function (PSF) of the star was slightly smaller than the slit width. This led to a slightly higher than nominal wavelength resolution achieved in this night \citep[super-resolution, e.g.][]{Nortmann2025,BoldtChristmas2025}. We found that the FWHM of the stellar PSF in spacial direction varied between 3 and 2.6 pixels, translating to a wavelength resolution between 100,000 and 117,000. A cross-correlation of the stellar spectra with a telluric template did not indicate a drift of the stellar PSF position in the slit, that is, no drift of the wavelength solution over time. No super-resolution was observed in the other nights. More details are given in \citet{Nortmann2025}.
\begin{table*}[ht]\renewcommand{\arraystretch}{1.5}
\centering
 \caption[]{Summary of the system parameters for the \mbox{GJ\,1214} system used in this study.}\label{tab:gj1214parameters}
\begin{tabular}{lllll}
 \hline \hline
  Adopted parameter & Symbol &   Value &Ref.
 \\ \hline
 \textit{Planet}&   &   & \\
Right Ascension& RA (hh:mm:ss) & 17:15:18.934 &(1) \\
Declination& Dec (dd:mm:ss)   & +04:57:50.067& (1) \\
Radius  &  $R_\mathrm{p}$ ($R_\mathrm{Earth}$) &  $2.733^{+ 0.052}_{-0.050}$& (2) \\
Mass  & $M_\mathrm{p}$ ($M_\mathrm{Earth}$) &  $8.17 \pm 0.43$ & (2)\\
Orbital period  &  $P_\mathrm{orb}$ (days) & $1.58040433(13)$&(2)\\
Orbital inclination & $i$ ($^\circ$) & $88.7^{+0.1}_{-0.1}$&(2)\\
Scaled semi-major axis &$a/R_\mathrm{s}$& $14.85\pm 0.16$&(2)\\
Time of mid-transit &$T_0$ (BJD$_\mathrm{TBD}$) & $2455701.413328^{+0.000066}_{-0.000059}$ &(2)\\
Surface gravity &$g_\mathrm{p}$ (m\,s$^{-2}$) &$10.65^{+0.67}_{-0.71}$&(2)\\
Night-side temperature& $T_\mathrm{night~side}$ (K) &$437 \pm 19$& (3) \\ 
Day-side temperature& $T_\mathrm{day~side}$ (K) &$553 \pm 9$ &(3)&\\ 
RV semi-amplitude &$K_\mathrm{p}$ & $102.5 \pm2$&(4) \\
\hline
 \textit{Star}&   &   & \\
Radius &$R_\star$ ($R_\odot$) & $  0.215\pm 0.008$&(2)\\
Mass& $M_\star$ ($M_\odot$) &  $0.178 \pm0.010$ &(2) \\
System velocity &$v_\mathrm{sys}$ (km\,s$^{-1}$) & +21.1 $\pm$ 1.0& (5, 6)\\
RV semi-amplitude &$K_\star$ (m\,s$^{-1}$) &14.36 $\pm$ 0.53 & (2) \\ 
\hline
\end{tabular}
\tablebib{1. \cite{Wenger2000}, 2. \cite{Cloutier2021}, 3. \cite{Kempton2023}, 4. Calculated from $K_\mathrm{p}=(2 \pi~ a~ \sin(i)/P)$, 5. \cite{Charbonneau2009}, 6. \cite{Kasper2020} (the latter  work corrects the sign of the value given in 5.)}
\end{table*}
\subsection{Determination of the optimal SYSREM iteration} \label{Sec:apendixsysremiterations}
\begin{figure}
\centering
\includegraphics[width=1\hsize]{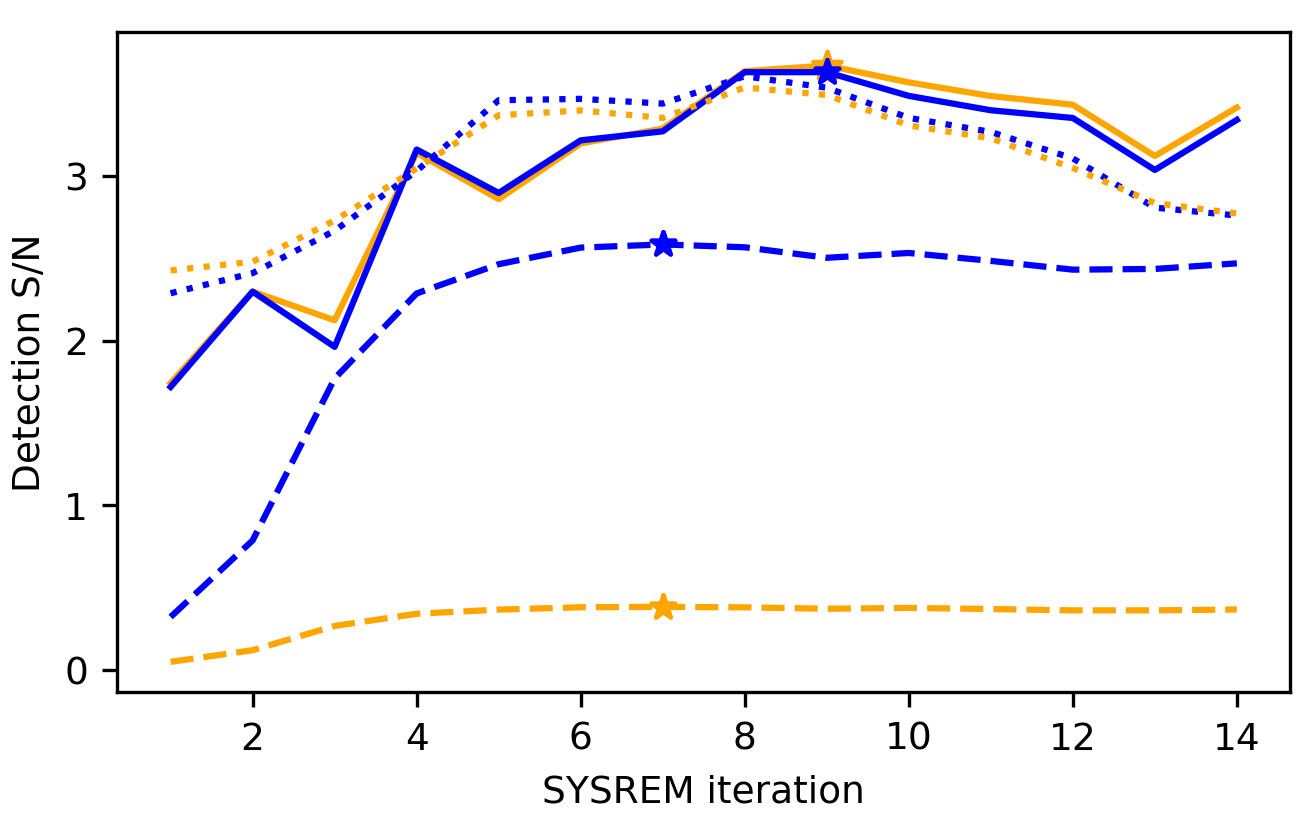}
  \caption{Behaviour of the cross-correlation peak for two models for the CO$_2$ signal in the real data in a 10 km s$^{-1}$ window around the expected planet signal (solid lines), for the injected template model (dashed lines) and for spurious peaks in the $K_\mathrm{p}$ - $v$ map outside of the vicinity of the expected planetary signal (dotted lines). Both models are calculated for  a metallicity of [M/H]=1.5; in blue we show the results for a model without clouds and in orange for a model with very high clouds at $\log_{10}(P_\mathrm{c})=-5.5$. The stars indicate the position at which the signal was measured with the highest S/N.}
     \label{fig:best_iteration_injectionrecovery}
\end{figure}

The two common approaches to find the best \texttt{SYSREM} iterations are to either inject a synthetic model into the data prior to the normalisation and find the iteration at which it can be recovered at the highest S/N \citep{Cheverall2023, Nortmann2025} or, the more model agnostic approach, to identify the iteration after which the change in the standard deviation of the residuals from one iteration to the next reaches a plateau \citep[e.g.][]{Herman2020, Herman2022, Deibert2021, RiddenHarper2023, Lesjak2025b}. Keeping the number of \texttt{SYSREM} iterations low is considered desirable as progressing iterations may start to remove the planetary signal. \\
It has, however, been shown that the common methods can underestimate the number of iterations needed to remove correlated residuals \citep{Lesjak2025b}. This may be due to the fact that both methods consider the noise as Gaussian by calculating it as the standard deviation (of the CCF in the first approach and of the data in the second approach). Both methods therefore do not accurately capture the effects of correlated residual noise which in turn may give rise to spurious peaks $>1\sigma$ in the $K_\mathrm{p}$ - $v$ maps. \\
In our case both methods suggested seven \texttt{SYSREM} iterations to be sufficient but the injection recovery method did not indicate a strong negative impact on the planet signal when increasing iterations slightly beyond that. When investigating the CCF results of the 
CO$_2$ analysis, we found that it benefits from two further iterations beyond the seventh, as these reduced the amplitude of spurious signals in the overall $K_\mathrm{p}$ - $v$ map while not also reducing a candidate planetary signal. We therefore decided to proceed with the ninth \texttt{SYSREM} iteration. This does not affect the overall results of the presence or absence of signals reported in this work. We plotted the behaviour of the peak CO$_2$ signal in the central part of the $K_\mathrm{p}$ - $v$ map (e.g. $K_\mathrm{p}=102.5\pm10$ km s$^{-1}$ and $v=0\pm10$ km s$^{-1}$), the amplitude of the highest noise peak in the region outside of the central part of the $K_\mathrm{p}$ - $v$ map and the peak signal of the 'injected and recovered' model as a function of \texttt{SYSREM} iterations in Fig. \ref{fig:best_iteration_injectionrecovery}.

\begin{figure}
\centering
\includegraphics[width=1\hsize]{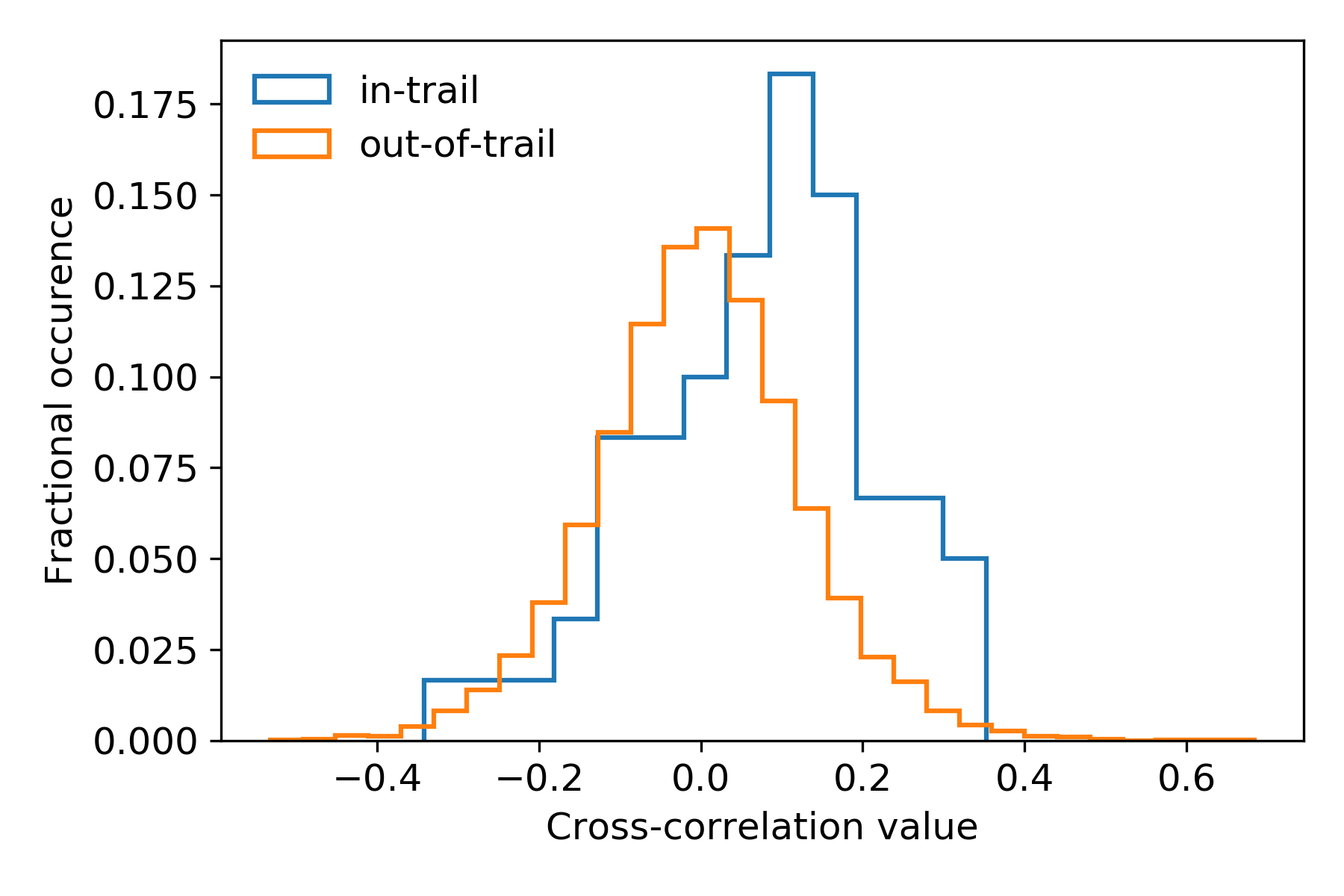}
  \caption{Histograms of the CCF values in- and out-of-trail. The ‘in-trail’ distribution corresponds to a 1 km s$^{-1}$ wide bin around $v=-1$~km~s$^{-1}$ for a CCF map aligned to the planet rest frame using $K_\mathrm{p}$ = 101 km s$^{-1}$.  The 'out-of-trail' distribution corresponds to the values obtained in the bins spanning from -250 to -10 and +10 to +250 km s$^{-1}$. Both distributions are obtained for the data obtained between the second and third contact of the planet transit.} 
     \label{fig:inouttrail}
\end{figure}
\begin{figure*}
\includegraphics[width=1\hsize]{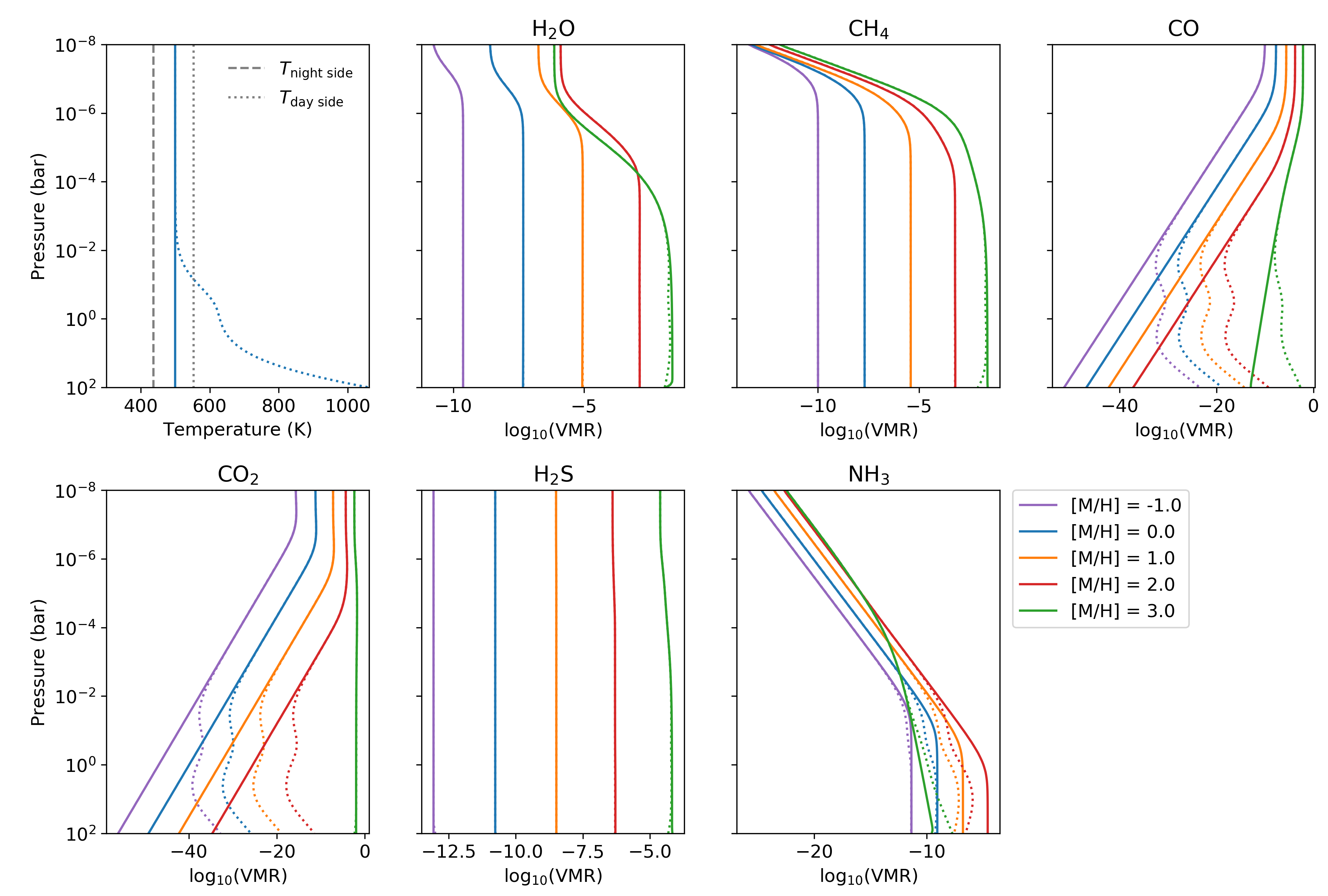}
  \caption{Pressure dependent temperature and abundance profiles. Top left panel: The isothermal temperature-pressure profile used to calculate the synthetic models used for the cross-correlation search for signals (solid blue line) compared to an example profile with physically expected increased temperatures at higher pressures (dotted blue line). The measured temperatures of the planet's day and night sides are indicated in dashed and dotted grey lines. Remaining panels: The volume mixing ratios (VMRs) of the individual molecules calculated from these $T$-$p$ profiles are shown for different assumed atmospheric metallicities [M/H] at constant, solar C/O=0.55 ratio.}
     \label{fig:template_tpprofile_masfunct}
\end{figure*}
\subsection{Model-filtering in the Bayesian retrieval}\label{sec:modelfiltering}
We use model pre-processing (also called filtering) during the Bayesian retrieval to recreate the distortions \texttt{SYSREM} imposed on the real data in the model. For this we use the approach suggested by \citet{Gibson2022}, with small tweaks to adjust for differences in our respective data analysis.  To filter the model we needed to bring it into the same format as the data, that is, for each of the 18 wavelength segments a 2D matrix with the two dimensions being time steps (number of exposures in the respective night) and wavelength channels (pixels) in the segment. The model in each exposure was Doppler shifted to correspond to the planet velocity based on the orbital parameters that were either free parameters of the retrieval or kept fixed. During the original \texttt{SYSREM} correction of the real data we saved the time-dependent vectors $u_n$ which were iteratively identified in the $n$ \texttt{SYSREM} iterations and are of the length of the number of spectra. We save them into the matrix $\mathbf{U}$, where $\mathbf{U}$ has the dimensions $L$ $\times$ $N$, with $L$ the number of spectra and $N$ the number of \texttt{SYSREM} iterations. We then calculated the matrix,
\begin{align}
\mathbf{F}=\mathbf{U} \big(\mathbf{\Lambda}  \mathbf{U} \big)^{\dagger} \mathbf{\Lambda.} 
\end{align}
where $\mathbf{\Lambda}$ is a diagonal matrix of 1/$\sigma_{\mathrm{av}}$ terms, with $\sigma_{\mathrm{av}}$ the mean of the uncertainties over wavelength. Deviating from \citet{Gibson2022} we did not add an extra column, $u_0$, containing ones as they did to account for their division by a master spectrum prior to \texttt{SYSREM}, a step not present in our analysis. The correction that needed to be applied to filter the model was then calculated in every step of the retrieval as,
\begin{align}
\mathbf{C}=\mathbf{F}\mathbf{M}=\mathbf{U} \big( \mathbf{\Lambda}  \mathbf{U} \big)^{\dagger} \big(\mathbf{\Lambda}  \mathbf{M}\big),
\end{align}
where $\mathbf{M}$ represents the 2D  matrix of the normalised model spectrum subtracted by 1. 
Since we subtracted each \texttt{SYSREM} iteration during the iteration process in which we obtained the cumulative correction matrix, but then divided instead of subtracted this correction matrix from the data, we also divide our model by our correction term to obtain the filtered model matrix
\begin{align}
\mathbf{M'}=(\mathbf{M}+1)/(\mathbf{C}+1)-1.
\end{align}
We note that applying the correction by subtraction, e.g. \mbox{$\mathbf{M'}= \mathbf{M}-\mathbf{C}$} instead, as was done by \citet{Gibson2022} for their version of the analysis, only leads to  negligible differences in the results, which are two orders of magnitude smaller than the model changes to which our retrieval is sensitive to. We tested this filtering against different injected models at several iterations and found it to reproduce the distortions reliably.
\begin{table*}\renewcommand{\arraystretch}{1.5}
\caption{CCF signal values for $K_\mathrm{p}$ - $v$ maps displayed in Fig. \ref{fig:model-ccf-kp}}
\begin{tabular}{c l c c c}  
\hline\hline  
Molecule & Peak S/N position & Value &Value at [$K_\mathrm{p,CO_2}$=101, $v_\mathrm{CO_2}=-1$] & Value at [$K_\mathrm{p}=102.5$, $v=0$]   \\ 
\hline  
H$_2$O &[$K_\mathrm{p}=100$, ~$v=-5$] &1.63  & 1.47 & 1.40 \\  
CH$_4$ &[$K_\mathrm{p}=113$, $v=+7$]& 0.95   &  0.40 & 0.27  \\  
CO  &[$K_\mathrm{p}=113$, $v=-1$] &1.84  &  1.56 & 0.71  \\  
CO$_2$ &[$K_\mathrm{p}=101$, $v=-1$]& 3.62 &  3.62  & 2.55 \\   
H$_2$S &[$K_\mathrm{p}=~93$, $v=-10$] &0.64 & $-1.73$ & $-0.97$ \\  
NH$_3$ &[$K_\mathrm{p}=~93$, $v= +$0]& 0.27   & $-0.12$& 0.08\\ 
\hline  
  \label{tab:CCFpeaks}
\end{tabular}
\tablefoot{CCF peaks and values in the vicinity of the expected planet signal e.g. maximal S/N reached within the range of $K_\mathrm{p}=102.5\pm10$ km s$^{-1}$ and $v=0\pm10$ km s$^{-1}$, and the precise S/N values reached at the position of the peak CO$_2$ signal  [$K_\mathrm{p}$=101, $v=-1$] as well as the position of the literature expected planet signal.}
\end{table*}
\begin{figure*}
\includegraphics[width=1\hsize]{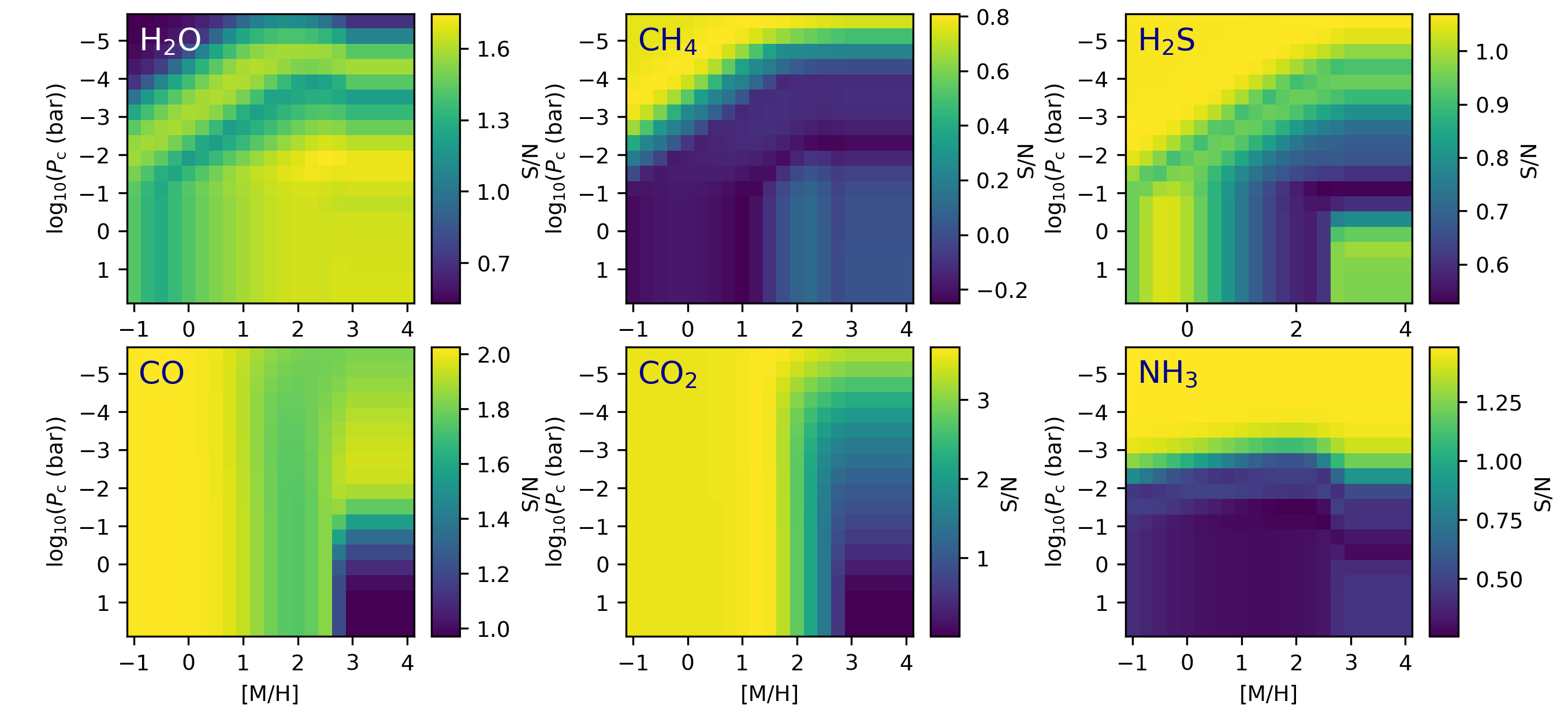}
  \caption{Peak signal in normalised $K_\mathrm{p}$ - $v$ maps computed from the real data with a grid of different model templates with varying metallicities [M/H] and opacity deck pressures, $P_\mathrm{c}$. The maximal S/N value in the vicinity of the expected planet signal, that is, the range of \mbox{$K_\mathrm{p}=102.5\pm10$~km~s$^{-1}$} and $v=0\pm10$~km~s$^{-1}$, was adopted for each explored model template. }
     \label{fig:detectiongrid}
\end{figure*}
\subsection{In- and out-of-trail test for the CO$_2$ signal} \label{sec:inouttrail}
The CCF signal should follow the planet velocity and only occur during transit. In the case of strong signals a trace can be seen in the phase resolved CCF-maps (middle column of Fig. \ref{fig:model-ccf-kp}). To test the robustness of signals in previous studies, the distributions of CCF points inside and outside of the planet trail were compared \citep{Birkby2013,Birkby2017,AlonsoFloriano2019}. Using the peak signal location we define our planet trail as a 1~km~s$^{-1}$  wide bin around $v=-1$~km~s$^{-1}$ offset from the expected rest frame calculated with $K_\mathrm{p}=101$~km~s$^{-1}$. We compare the distribution of the 'in-trail' CCF values between the second and third contact of the transit with the distributions of 'out-of-trail' values obtained during transit. We define the out of trail regions as those spanning $-250$~km~s$^{-1}$  to $-10$~km~s$^{-1}$ and +10~km~s$^{-1}$  to 250~km~s$^{-1}$.  The results are shown in Fig.~\ref{fig:inouttrail}. The distribution of CCF values obtained in-trail is significantly offset from the out-of-trail distribution, which scatters around zero. Fitting Gaussian profiles to the data yields midpoints of $0.068 \pm 0.019$ for the 'in-trail'  and $-0.001 \pm 0.001$ for the 'out-of-trail' distribution. A Welch $t$-test rejects the hypothesis that the in-trail and out-of-trail distributions are drawn from the same parent population at a $3.4\sigma$ confidence. We excluded nights 1 and 2 in these tests due to the heavy telluric masking and consequent lack of CCF signals (see Sect. \ref{sec:signaldecomposition}), even though including them still yielded the same conclusions.
\begin{figure}
\includegraphics[width=0.75\hsize]{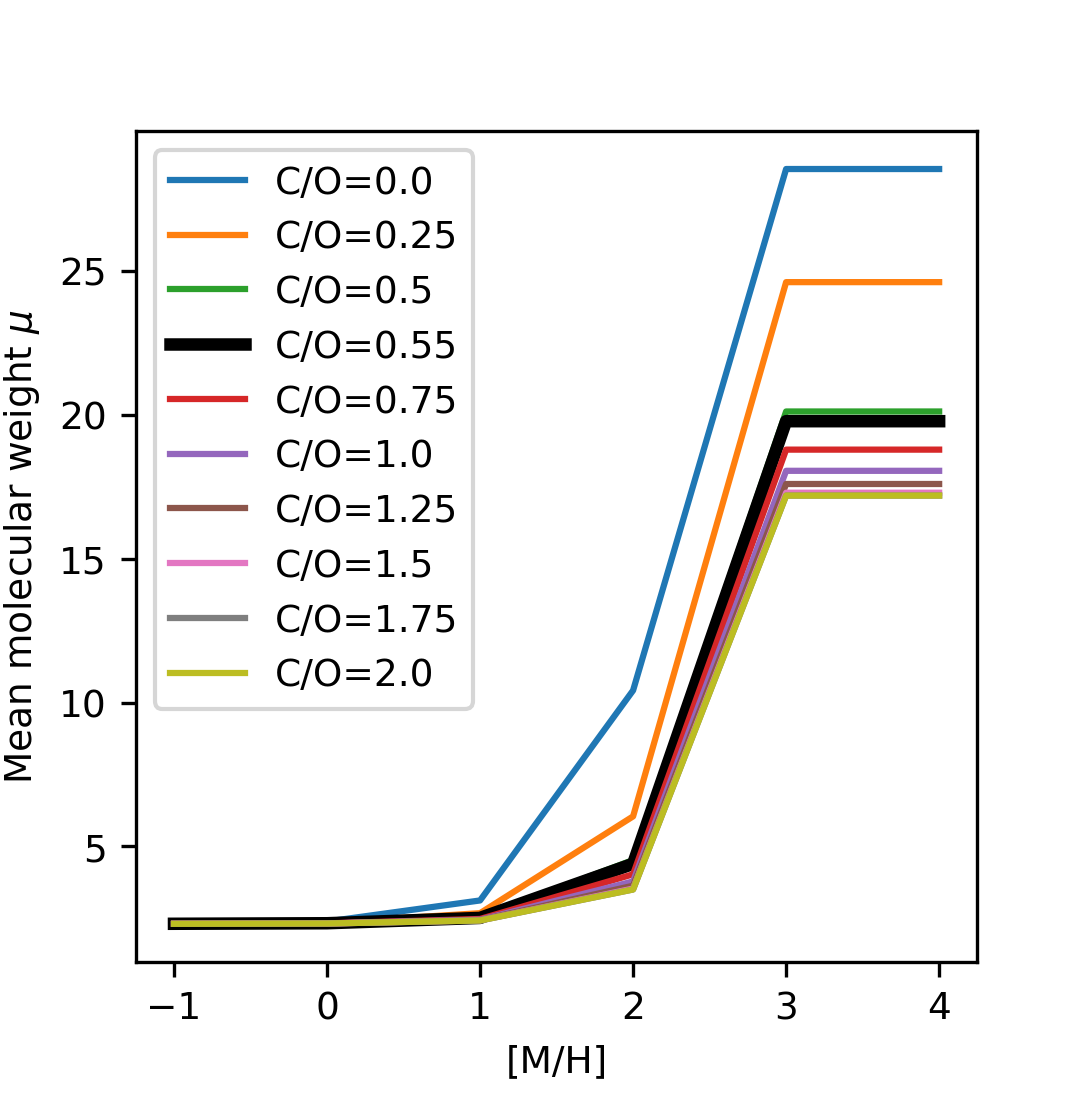}
  \caption{Change of the mean molecular weight, $\mu$, with increasing metallicity of the atmosphere. The different coloured curves show the behaviour for different assumed C/O ratios. The solar C/O ratio assumed in the injection recovery grids is indicated in bold black. }
     \label{fig:mu}
\end{figure}
\begin{figure}[h!]
\includegraphics[width=1\hsize]{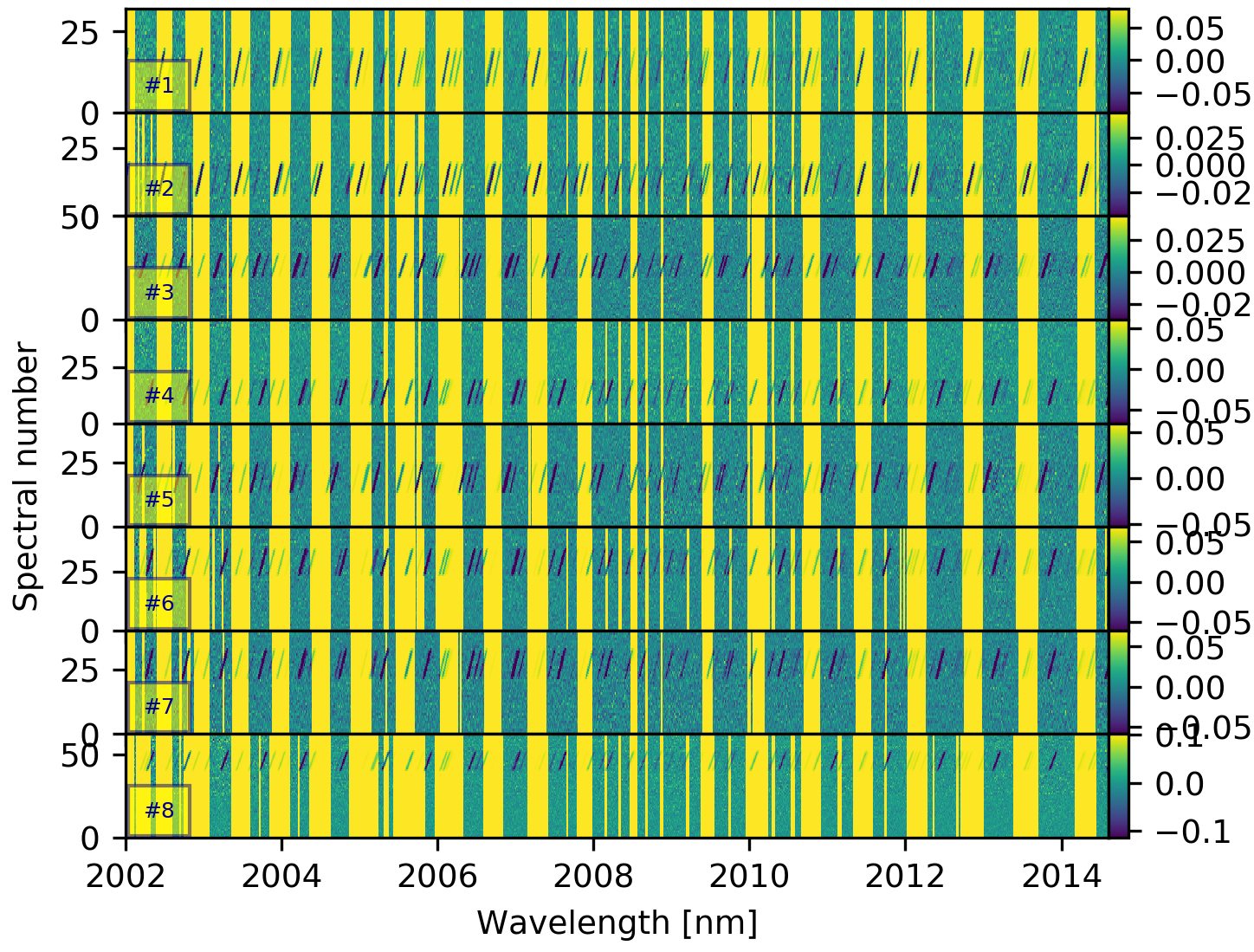}
  \caption{Position of the planet signal with respect to the masked telluric regions in the individual nights. Shown is the final stage of the data analysis with a CO$_2$ planet signal injected at 200 times the strength expected for GJ\,1214\,b. The masked regions were set to zero (shown in yellow), to allow visualisation of the amount of planet signal falling into the masked regions. Results are shown on the example of the third wavelength segment (2002.0-2014.6 nm), with strong CO$_2$ absorption. This is the same as is displayed in the bottom panel of Fig. \ref{fig:analysis} for night 5.   
  }
   \label{fig:analysis_alln}
\end{figure}
\begin{figure*}[h!]
\centering
\includegraphics[width=1\hsize]{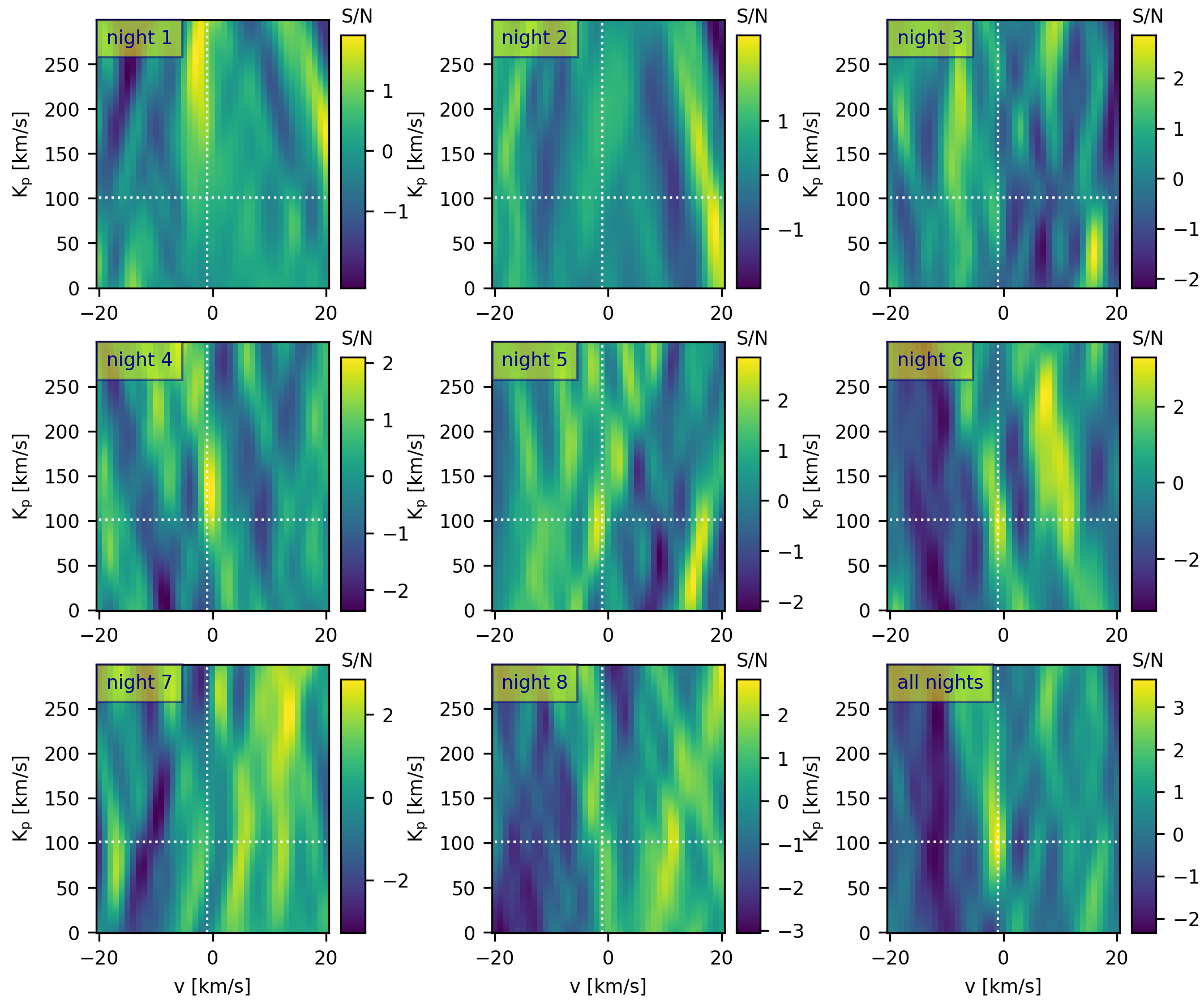}
  \caption{$K_\mathrm{p}$-$v$ maps around the location of the peak signal for CO$_2$ for the eight individual nights. The white dotted lines indicate the location of the peak signal in the combined map, $K_\mathrm{p}=101$~km~s$^{-1}$ and $v=-1$~km~s$^{-1}$.}
   \label{fig:kpmaps_allnights}
\end{figure*}

\subsection{Simulation of ELT-data for injection recovery-tests} \label{sec:ELTsims}
To complement the simulated constraints we would be able to place on GJ\,1214\,b with 16 nights of VLT/CRIRES$^+$ observations, we simulated the expected constraints on the planet atmosphere that could be derived with a hypothetical K-band spectrograph mounted at ESO's Extremely Large Telescope (ELT) currently under construction in Chile. 
While such a spectrograph is considered as part of the ANDES project \citep{Maiolino2013}, its specifications remain undetermined \citep{Palle2025}. We thus opted to model a hypothetical ELT/CRIRES$^+$ spectrograph, with identical detector format, echelle ordering, wavelength coverage, spectral resolution, throughput, and overheads, only instead with S/N scaled from an 8.2 m telescope to a 39 m telescope, assumed here to be a factor of 22.6.  
To limit the comparison to just the increased S/N enabled by the ELT and avoid confounding factors that would be introduced by, for example, sampling a different set of barycentric velocities, we simulated the transit events of our CRIRES$^{+}$ observing nights 3 to 8. We omitted simulations of nights 1 and 2 due to the unfavourable total velocity shift between the planet lines and telluric lines in these nights (see Fig.~\ref{fig:obs_stats} and Sect.~\ref{sec:signaldecomposition}). Our transmission spectroscopy simulation methodology is described in more detail in \citet{BoldtChristmas2025} and \citet{Piskunov2025}, but we present a short summary as follows here. We used a stellar spectrum of GJ\,1214 computed from the grid of MARCS \citep{Gustafsson2008} model spectra first described by \citet{Nordlander2019} and using the stellar parameters from \citet{Cloutier2021}. Our telluric spectra were generated from the K band per-molecule (H$_2$O, CH$_4$, N$_2$O, CO$_2$, CO) template transmission spectra computed by \citet{Koehler2025} for the \texttt{viper} software package, and we allowed H$_2$O absorption and slit losses to smoothly vary throughout each night to simulate changes in humidity and seeing respectively. Using our  CRIRES$^+$ observations as a reference, that is peak S/N $\sim$ 60-80 per 240 second exposure, we scaled the total expected counts by 22.6$\times$ and halved the exposure times to reduce risk of saturation, resulting in a peak S/N of $\sim$ 270 per 120 second ELT/CRIRES$^+$ exposure. The planet spectra were injected into the data and recovered as described in Sec. \ref{sec:injrecov}. We show the resulting constraints for one observing night with the ELT and six cumulative nights in Fig. \ref{fig:futuresim}.
\begin{figure*}
\includegraphics[width=1\hsize]{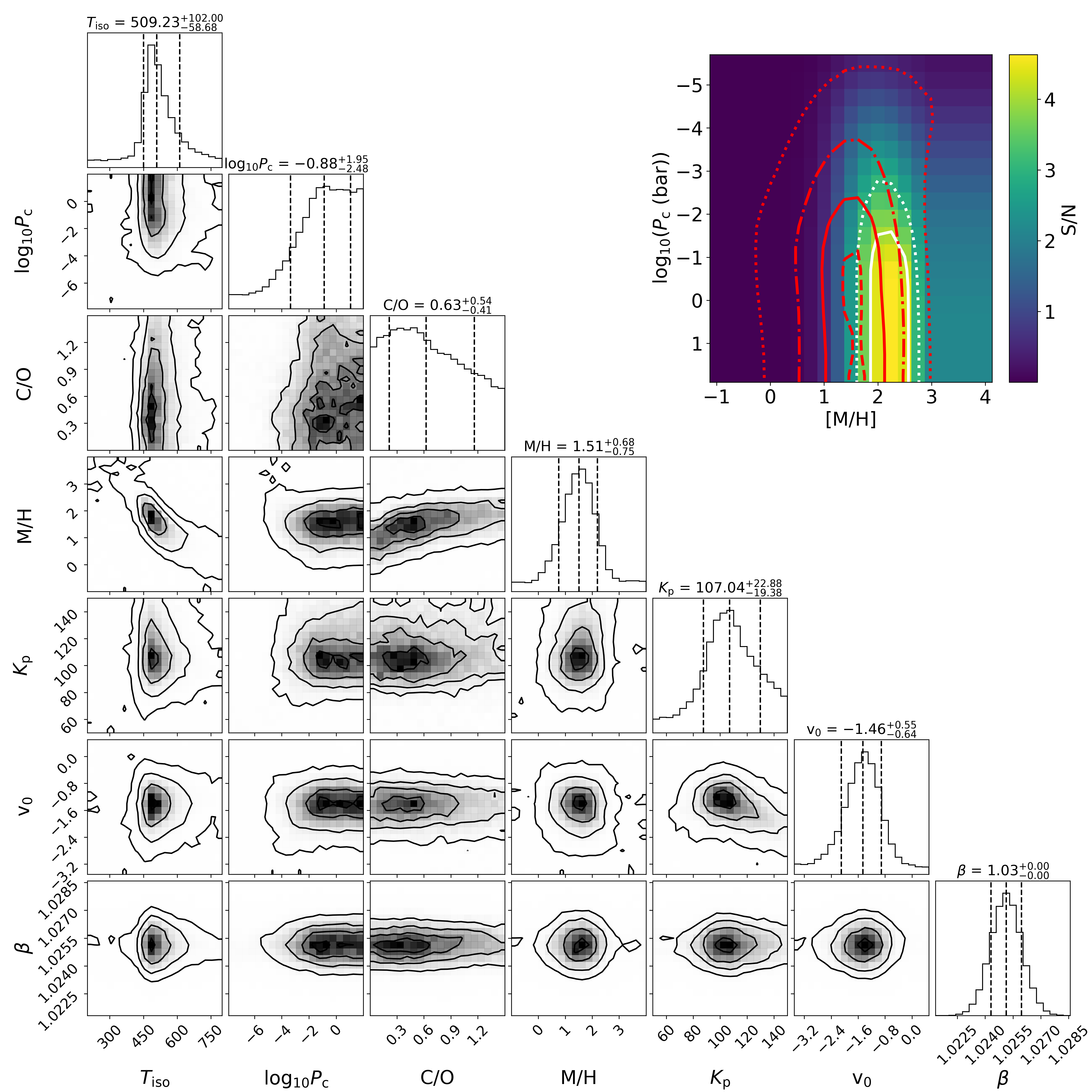}
  \caption{
  Results of the Bayesian retrieval run on our GJ\,1214\,b high-resolution data. 
   The free parameters of the retrieval were the planet's temperature, $T_\mathrm{iso}$, used to calculate an isothermal $T$-$p$ profile, the opacity deck top pressure, $\log_{10}(P_\mathrm{c}$), the C/O ratio and metallicity [M/H] of the atmosphere, the system parameters $K_\mathrm{p}$ and $v_0$,  and a scaling factor for the data uncertainties, $\beta$. Upper right: Posterior distributions for the metallicity [M/H] and opacity deck pressure, $P_\mathrm{c}$,  plotted in red as contours at the 0.5, 1.0, 1.5, and 2.0 $\sigma$ level on top of the grid of expected S/N for different [M/H] and $P_\mathrm{c}$ from injection-recovery tests, for which S/N=3 and 4 contours are indicated as white lines.}
     \label{fig:retrivalresults}
\end{figure*}

\begin{figure*}
\includegraphics[width=1\hsize]{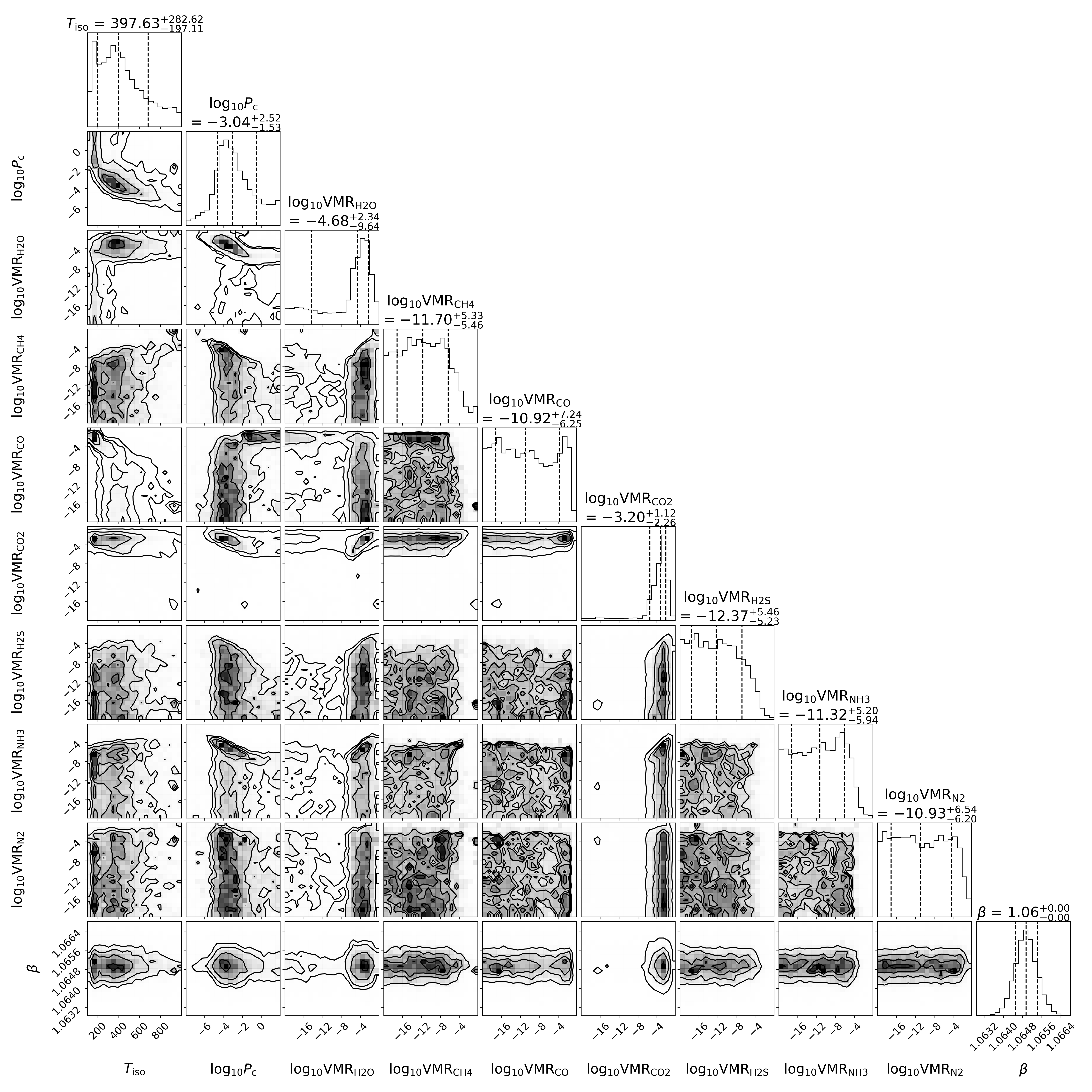}
  \caption{
  Results of the Bayesian retrieval run on our GJ\,1214\,b high-resolution data. 
   The free parameters of the retrieval were the planet's temperature, $T_\mathrm{iso}$, used to calculate an isothermal $T$-$p$ profile, the opacity deck top pressure, $\log_{10}(P_\mathrm{c}$), the volume mixing ratios (VMRs) of 7 molecules,  and a scaling factor for the data uncertainties, $\beta$.}
     \label{fig:retrivalresults_freechem}
\end{figure*}

\begin{figure*}
\includegraphics[width=1\hsize]{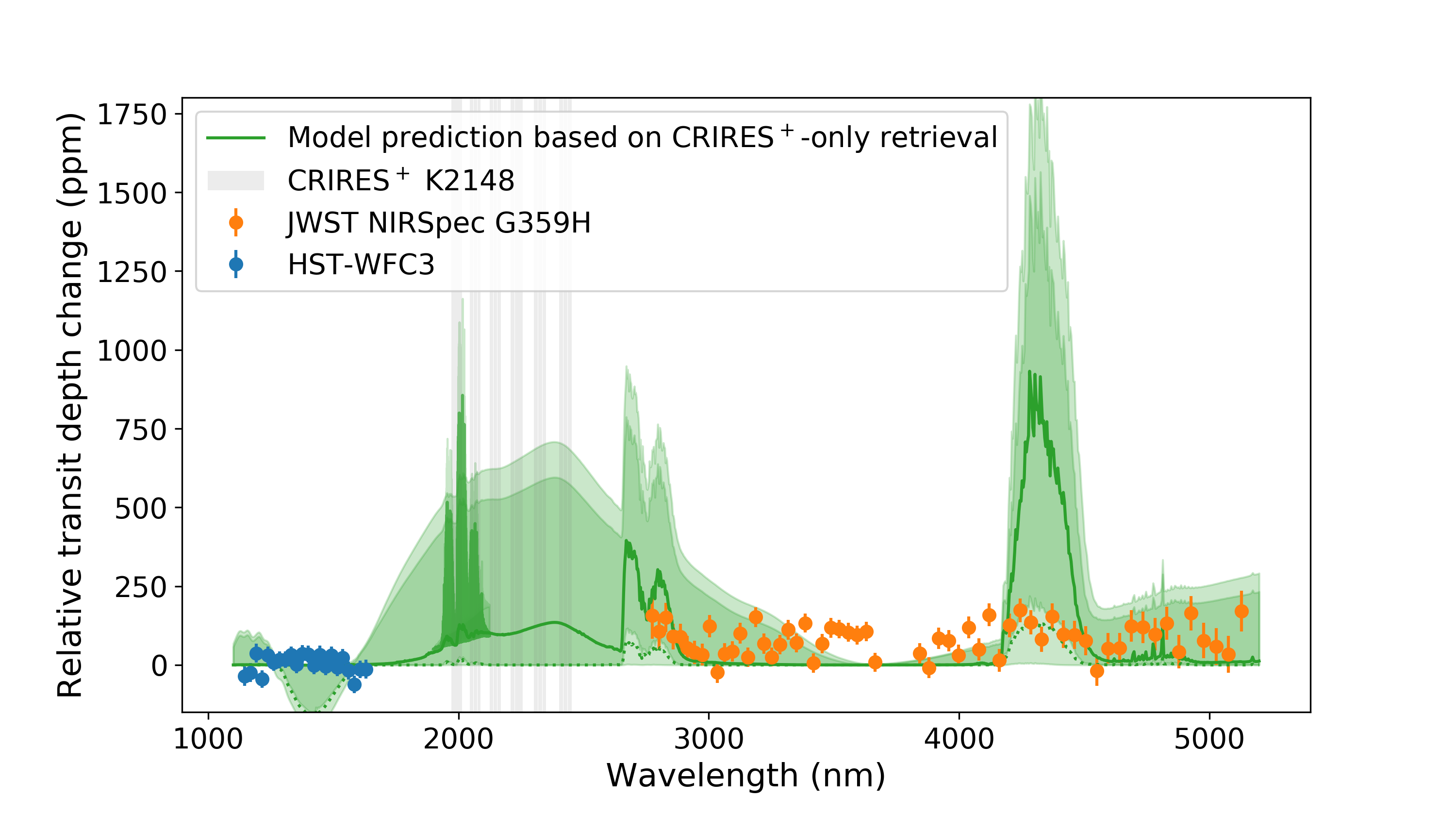}
  \caption{Predicted relative transit depth change from the equilibrium chemistry CO$_2$-only model retrieved for CRIRES$^+$ data  (green). The model predictions are plotted in high-resolution in the wavelength region overlapping with the CRIRES$^+$ measurements indicated by the grey shaded areas and in low-resolution for the rest of the wavelength range. Space-based measurements from HST (blue) and JWST (orange) are also shown. The 1$\sigma$ and 1.5$\sigma$ uncertainties of the model are indicated as darker and lighter green shaded areas, respectively. We also mark the lower 1.1$\sigma$ uncertainty boundary with a dotted green line as a potential match to the amplitude of the CO$_2$ bump in the JWST measurements at 4200 nm. The HST-WFC3 and JWST NIRSpec G359H measurements were taken from the online data provided by \citet{Schlawin2024} containing the HST data adapted from \cite{Kreidberg2014}.
  }
     \label{fig:retmodel_vs_jwst} 
\end{figure*}

\end{document}